\documentclass[aps,article,superscriptaddress,nofootinbib]{revtex4-2}

\usepackage[utf8]{inputenc}
\usepackage[fleqn]{amsmath}
\usepackage{amssymb}
\usepackage{graphicx}
\usepackage{xcolor}
\usepackage{epstopdf, epsfig}
\usepackage{graphicx}
\usepackage{mathtools}
\usepackage{bm}
\usepackage{soul}

\newcommand{\vz}{{\bf z}}

\newcommand{\vE}{{\bf E}}
\newcommand{\vB}{{\bf B}}

\newcommand{\Wcs}{W_{\rm\scriptscriptstyle{CS}}}
\newcommand{\Wbw}{W_{\rm\scriptscriptstyle{BW}}}

\begin{document}

\title{Multiplicity of electron- and photon-seeded electromagnetic showers 
\\at multi-petawatt laser facilities}

\author{M. Pouyez}
\thanks{\url{mattys.pouyez@polytechnique.edu}}
\affiliation{LULI, Sorbonne Université, CNRS, CEA, École Polytechnique, Institut Polytechnique de Paris, F-75255 Paris, France}
\author{A.~A. Mironov}
\affiliation{LULI, Sorbonne Université, CNRS, CEA, École Polytechnique, Institut Polytechnique de Paris, F-75255 Paris, France}
\author{T. Grismayer}
\affiliation{GoLP/Instituto de Plasmas e Fusão Nuclear, Instituto Superior Técnico, Universidade de Lisboa, 1049-001 Lisboa, Portugal}
\author{A. Mercuri-Baron}
\affiliation{LULI, Sorbonne Université, CNRS, CEA, École Polytechnique, Institut Polytechnique de Paris, F-75255 Paris, France}
\affiliation{GoLP/Instituto de Plasmas e Fusão Nuclear, Instituto Superior Técnico, Universidade de Lisboa, 1049-001 Lisboa, Portugal}
\author{F. Perez}
\affiliation{LULI, CNRS, CEA, Sorbonne Universit\'{e}, École Polytechnique, Institut Polytechnique de Paris, F-91128 Palaiseau, France}
\author{M. Vranic}
\affiliation{GoLP/Instituto de Plasmas e Fusão Nuclear, Instituto Superior Técnico, Universidade de Lisboa, 1049-001 Lisboa, Portugal}
\author{C. Riconda}
\affiliation{LULI, Sorbonne Université, CNRS, CEA, École Polytechnique, Institut Polytechnique de Paris, F-75255 Paris, France}
\author{M. Grech}
\affiliation{LULI, CNRS, CEA, Sorbonne Universit\'{e}, École Polytechnique, Institut Polytechnique de Paris, F-91128 Palaiseau, France}

\begin{abstract}
    Electromagnetic showers developing from the collision of an ultra-intense laser pulse with a beam of high-energy electrons or photons are investigated under conditions relevant to future experiments on multi-petawatt laser facilities. A semi-analytical model is derived that predicts the shower multiplicity, i.e. the number of pairs produced per incident seed particle (electron or gamma photon). The model is benchmarked against particle-in-cell simulations and shown to be accurate over a wide range of seed particle energies (100 MeV - 40 GeV), laser relativistic field strengths ($10 < a_0 < 1000$), and quantum parameter $\chi_0$ (ranging from 1 to 40). It is shown that, for experiments expected in the next decade, only the first generations of pairs contribute to the shower while multiplicities larger than unity are predicted. Guidelines for forthcoming experiments are discussed considering laser facilities such as Apollon and ELI Beamlines. The difference between electron- and photon seeding and the influence of the laser pulse duration are investigated.
\end{abstract}
\maketitle
\section{Introduction}\label{sec:intro}

Since its first developments 90 years ago \cite{bhabha1937passage,carlson1937multiplicative,rossi1941cosmic}, quantum electrodynamics (QED) has been established as the best tested, most precise physical theory. Its subbranch, strong-field QED (SF-QED), which describes light-matter interaction in the presence of very strong electromagnetic fields, however, remains only partially tested. Over the last decade, it has become the focus of extensive research \cite{di2012extremely,fedotov2016quantum} as new laser and accelerator facilities worldwide allow us to put this theory to the test. 

Among the most exotic phenomena predicted by SF-QED is the (potentially copious) production of electron-positron pairs in so-called QED cascades developing in strong electromagnetic fields. At the heart of these cascades are two QED processes:  nonlinear Breit-Wheeler pair production~\cite{breit1934collision}, that is the decay of a photon into an electron-positron pair under the influence of a strong electromagnetic field, 
and nonlinear Compton scattering that is in the emission of a high-energy photon by an electron (or positron) subjected to a strong electromagnetic field~\cite{gauge_schwinger_1951,quantum_nikishov_1964,quantum_ritus_1979}.

These exotic processes are omnipresent in extreme astrophysical environments, such as neutron stars, pulsars, and magnetars \cite{pulsar_goldreich_1969,physics_harding_2006, plasma_uzdensky_2014}, where intense magnetic fields give rise to the copious emission of electron-positron pairs and high-energy $\gamma$ photons \cite{timedependent_timokhin_2010,pair_medin_2010}. Creating pair plasmas in the laboratory would be crucial to understanding 
the dynamics of highly magnetized environments and testing SF-QED. 
While it still poses a significant challenge \cite{generation_sarri_2015}, the advent of laser systems of unprecedented power may soon allow for the creation of QED cascades in the laboratory and the abundant generation of pairs in the laboratory.

Indeed, several multi-petawatt laser facilities are being developed worldwide~\cite{generation_bahk_2004,petawatt_danson_2019}:
Apollon~\cite{apollon_papadopoulos_2016},
CoReLS~\cite{nam2018performance}, 
ELI~\cite{ELI},
EP-OPAL~\cite{technology_bromage_2019},
Vulcan~\cite{vulcan_hernandezgomez_2010} 
or ZEUS~\cite{ZEUS}.
These facilities will allow to deliver ultra-short (20 fs - 150 fs) light pulses with intensities beyond $10^{23}$ W/cm$^2$ \cite{realization_yoon_2021}. 
Various approaches have been proposed to use such light pulses to drive abundant pair production. 
These include irradiating a solid target with an ultra-intense light pulse~\cite{relativistic_chen_2009,dense_ridgers_2012,high_liang_2015,dense_zhu_2016,_zi_2023}, 
colliding two (to several) laser pulses to construct electromagnetic fields structures (e.g. rotating field) maximizing pair production~\cite{possibility_bell_2008,pair_kirk_2009,limitations_fedotov_2010,qed_elkina_2011,optimized_gelfer_2015,laser_grismayer_2016,qed_jirka_2017,electronpositron_vranic_2017,gonoskov2017ultrabright}, 
colliding a laser pulse with a beam of relativistic electrons or photons~\cite{burke1997positron,generation_mironov_2016,lobet2017generation,blackburn2017scaling,mercuri2021impact,golub2022nonlinear}, 
among others~\cite{creation_yu_2019,creation_martinez_2023}.

Creating a dense plasma of pairs in the laboratory would most certainly require generating a particular type of QED cascade, 
called an avalanche, in which the number of pairs increases exponentially with time \cite{seeded_grismayer_2017,qed_luo_2018,mercuri-baron_2024}. Entering the avalanche regime however often requires the use of several ultra-intense laser beams synchronized in space and time to build up electromagnetic field configurations allowing for particle re-acceleration and thus sustaining the avalanche process.
This renders the experiment currently inaccessible. 
An alternative, albeit less efficient, method involves colliding a beam of ultrarelativistic electrons 
\cite{blackburn2017scaling,lobet2017generation} or photons \cite{mercuri2021impact} against the intense laser pulse, leading to so-called electromagnetic showers. 
In this scenario, re-acceleration in the laser field has limited influence and the resulting shower (electron, positron, and photons) draws its energy from the initial (seed) electron or photon beam\cite{electromagnetic_bulanov_2013,collapse_mironov_2014}. 
To this day only the seminal SLAC E-144 experiment~\cite{burke1997positron} has demonstrated electron-positron production from the collision of an ultra-relativistic electron beam with a high-intensity laser pulse.  
However, due to the quite limited intensity of the laser employed in this experiment,  very few pairs were produced: $\sim 106$ pairs/21962 collisions.
Exploiting the capacities of multi-beam, multi-petawatt laser facilities will allow to enter the regime of abundant pair production, where any seed particle (electron or photon) injected into the focal volume of the ultra-intense laser pulse will lead to at least an electron-positron pair. 
This was for instance demonstrated through three-dimensional particle-in-cell simulations exploring conditions relevant to the Apollon laser facility for which it was shown that electron-seeded showers could lead to the production of high-density sources of ultra-relativistic electron-positron pairs accompanied by a brilliant flash of high-energy $\gamma$ photons.\\

In this paper, we explore both electron- and photon-seeded showers to predict a crucial observable in future experiments: the number of produced pairs.
More precisely, we derive a semi-analytical model that predicts the number of pairs produced per incident (seed) particle, 
henceforth referred to as the shower multiplicity. This model builds on previous works by Blackburn et al.~\cite{blackburn2017scaling} for electron-seeded showers and Mercuri-Baron et al.~\cite{mercuri2021impact}
 for photon-seeded ones. By alleviating some hypotheses made in these previous works, we significantly extend their domain validity to cover the region of parameters (seed-particle energy, laser field strength, pulse duration, etc) relevant to future experiments at multi-petawatt laser facilities\footnote{While the model can be generalized to any laser beam polarization and angle of incidence, the results are specifically presented for linear polarization and head-on collision.}. Throughout this paper, the model is tested against particle-in-cell (PIC) simulations performed with the open-source code Smilei~\cite{smilei_derouillat_2018} accounting for the relevant QED processes~\cite{monte_duclous_2011,contemporary_arber_2015,extended_gonoskov_2015,particle_vranic_2015,modeling_lobet_2016}. Doing so allows us to test the hypotheses of the model (see Sec.~\ref{sec:model:hyp} for details) as well as to get a deeper insight into the physics of electromagnetic showers.\\

The paper is structured as follows. In Sec.~\ref{sec:model} we describe a reduced model that predicts the number of pairs in a one-dimensional geometry. A key hypothesis introduced here is to assume that, under conditions relevant to our study, only the first generations of pairs contribute to the shower. The model is benchmarked against PIC simulations in Sec.~\ref{sec:pic1d}. This allows us to discuss the domain of validity of the model as well as to extract physical insight into the shower development and provide guidelines for future experiments. We discuss in detail the difference between photon and electron seeding, and that between short (few cycles) and long (150~fs) laser pulses. In Sec.~\ref{sec:3d}, we extend our predictions to a realistic three-dimensional case and test our extended model against PIC simulations. A full theoretical guide map for the shower multiplicity is derived and discussed as a function of the incident (seed) particle energy and laser field strength (intensity) focusing in particular on the cases of Apollon-like (Ti:Sapphire, 20~fs-long) and ELI-Beamlines-A4-Aton-like (Ne:Glass, 150~fs-long) laser pulses.
Finally, we summarize our conclusions in Sec.~\ref{sec:conclusion}.

\section{Reduced model}\label{sec:model}

SI units and standard notations for physical constants will be used throughout the paper:
$c$ stands for the speed of light in vacuum, $m$ for the electron mass, $e$ for the elementary charge, $\alpha = e^2/(4\pi\varepsilon_0\hbar c)$ for the fine structure constant, $\varepsilon_0$ for the permittivity of vacuum and $\tau_c = \hbar/(m c^2)$ for the Compton time.

\subsection{Objectives and hypotheses of the model}\label{sec:model:hyp}

\begin{figure}[t]
    \begin{center}
    \includegraphics[width=1.\linewidth]{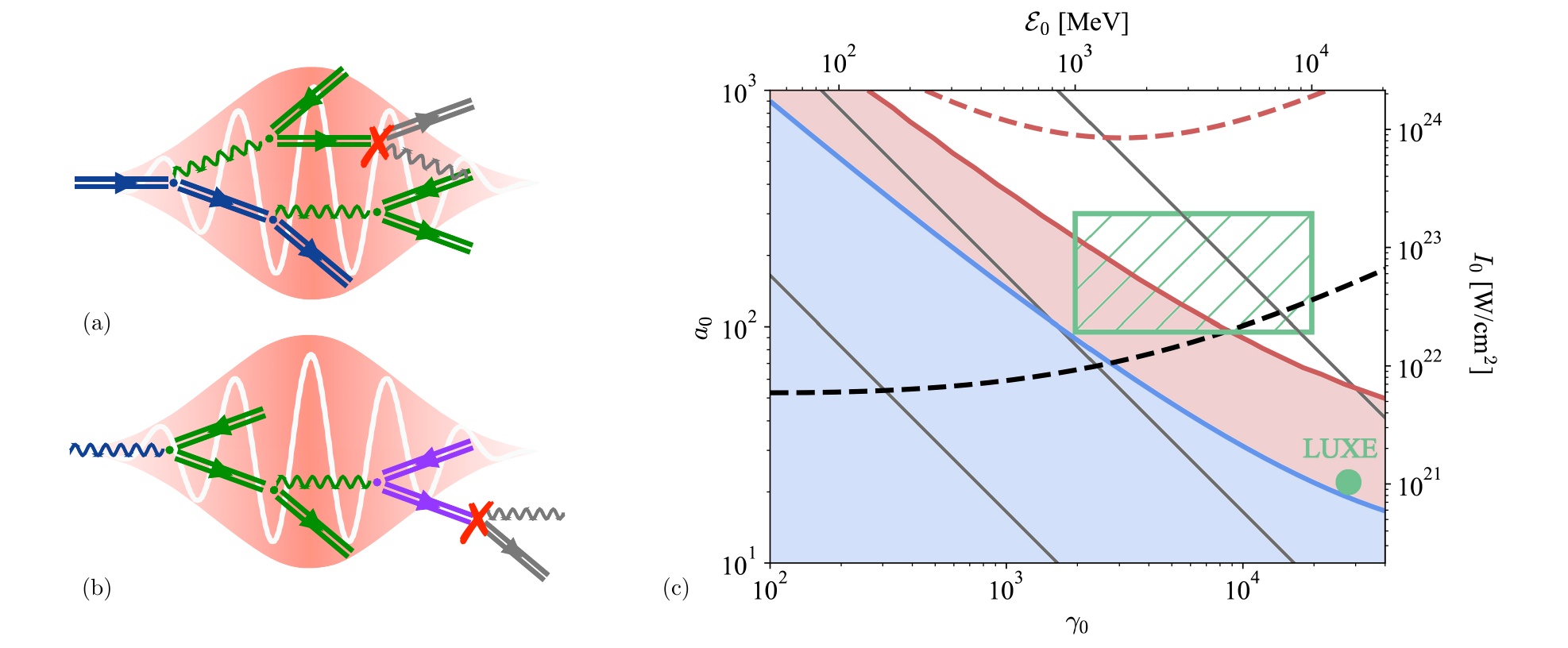}
    \end{center}
    \caption{Schematic of (a) electron-seeded and (b) photon-seeded electromagnetic showers showing the splitting of the process
    in successive generations. For electron-seeded showers, only the first generation of pairs (in green) is accounted for. For photon-seeded showers, the first (in green) and second (in purple) generations are accounted for. (c) Parameter-space studied in this work as a function of the seed-particle energy $\mathcal{E}_0 = \gamma_0 mc^2$ 
    and laser field strength $a_0=e E_0/(m c \omega)$ (correspondingly, intensity $I_0$). Particular attention is paid to the rectangular green region
    that corresponds to the operating regime of upcoming experiments on multi-petawatt laser facilities. 
    Here $\lambda=0.8~{\rm \mu m}$ and $\tau_p = 20~{\rm fs}$ were used.}\label{fig:intro}
\end{figure}

We present a reduced model for electromagnetic showers developing in the head-on collision of 
a high-energy electron or photon beam with an ultra-intense laser pulse,
see Fig.~\ref{fig:intro} (a) and (b).

The initial (seed) beam is considered to be a point-like (zero spatial extension and angular aperture) beam of initially monoenergetic particles 
with energy $\mathcal{E}_0 = \gamma_0\,mc^2$ for photons and $\mathcal{E}_0 = (\gamma_0-1)\,mc^2$ for charged particles. In this work, we will focus on ultra-high energy seed particles, $\gamma_0 \gg 1$, focusing
in particular on the range $\gamma_0 \in [100-40000]$, i.e. energies in the range 50 MeV - 20 GeV.
Furthermore, we will consider optical laser systems with a micrometric wavelength $\lambda$, and large
relativistic field strength parameter $a_0 = e E_0/(m c \omega) \gg 1$ (with $E_0$ the maximum laser electric field and 
$\omega=2\pi c/\lambda$ the laser angular frequency).
As our interest lies in future experiments on multi-petawatt laser systems, 
we focus on $a_0 \in [10,1000]$, 
corresponding to $I_0\lambda^2 \sim 10^{20} - 10^{24}~{\rm W/cm^2 \mu m^2}$ (with $I_0 = c\,\varepsilon_0 E_0^2/2$ the peak laser intensity).
Furthermore, predictions of the number of produced pairs will be discussed for two types of laser systems:
ultra-short (20 fs) Ti:Sapphire ($\lambda=0.8~{\rm \mu m}$) laser pulses as delivered at Apollon, CoReLS, or ZEUS, 
and longer (150 fs) Nd:Glass ($\lambda=1.057~{\rm \mu m}$) laser pulses as delivered by the L4 ATON laser beam at ELI Beamlines.\\

The range of parameters discussed in this work is reported in Fig.~\ref{fig:intro}(c) 
considering $\lambda=0.8~{\rm \mu m}$ and a full-width-at-half-maximum-intensity laser pulse duration $\tau_p=20~{\rm fs}$.
Superimposed is a rectangular region that corresponds to seed particle energies in between 1 and 10 GeV and field strengths $a_0$ in between $95$ and $300$ computed considering a 1PW or 10 PW Gaussian laser pulse
focused with an angular aperture $f/3$. It is relevant to most forthcoming experiments on multi-petawatt laser systems.

We also report the value of the maximum quantum parameter $\chi_0$ a particle (either photon, electron or positron)
with energy $\gamma_0\,mc^2$ can achieve colliding head-on with an electromagnetic wave of relativistic field strength $a_0$:
\begin{eqnarray}
    \label{eq:chi0}
    \chi_0 &\equiv& 2\,\frac{\hbar\omega}{m c^2}\,\gamma_0\,a_0 
    \simeq 0.485\,\left(\frac{1~{\rm \mu m}}{\lambda}\right)\,\left(\frac{\gamma_0}{1000}\right) \left(\frac{a_0}{100}\right)
    \simeq 0.812\,\left(\frac{\mathcal{E}_0}{1~{\rm GeV}}\right) \sqrt{\frac{I_0}{10^{22}~{\rm W/cm^2}}}\,.
\end{eqnarray}
In this representation, contours of iso-$\chi_0$ are oblique lines
and iso-contours $\chi_0 = 0.1, 1$ and $10$ are reported as grey lines.

We further report two regions of interest.
The first, blue one, corresponds to the region of validity of the model developed by Blackburn et al.~\cite{blackburn2017scaling}
for electron-seeded showers. It is limited to low $\chi_0$ values which is not of direct interest for
multi-petawatt laser facilities, but rather corresponds to regimes explored coupling standard accelerators with near-100 TW laser systems, 
as planned for experiments such as LUXE or FACET-II.
The second, in red, corresponds to the domain of validity of the model proposed by Mercuri-Baron et al.~\cite{mercuri2021impact}
for photon-seeded showers. More precisely, it corresponds to the limit of the so-called soft-shower regime in which pair production
is dominated by the decay of the seed photons only. This region covers part, but not all of the (green-hatched) region of interest 
for multi-petawatt laser systems. \\

Our objective is to extend the region of validity of both models to cover the whole region of interest for multi-petawatt laser systems.
To do so, we will go beyond some of the hypotheses made in the previous models.
To extend Blackburn's model to higher $\chi_0$, we (i) make the distinction between rates and probabilities, 
and (ii) include the contribution of intermediate-energy electrons to the shower.
To extend Mercuri-Baron's model ~\cite{mercuri2021impact}, we go beyond the soft-shower regime and account for the secondary pairs considering that primary pairs can radiate away
high-energy photons that can, in turn, decay into new electron-positron pairs [see Fig.~\ref{fig:intro}(b)].\\

By doing so, we derive a set of equations that can be integrated numerically.
To make this model tractable and gain insights into the development of the shower, 
we retain some hypotheses which we now detail.

First, we split all species (photons, electrons, and positrons) into successive generations which we treat separately, each species generation being described 
by its own distribution function (for a treatment of the full distribution 
functions see Appendix~\ref{appendixA}).
When considering electron-seeded showers [see Fig.~\ref{fig:intro}(a)], 
the seed electrons will emit a first generation of (primary) high-energy photons that may then decay in a first generation of (primary) electron-positron pairs.
Our model of electron-seeded showers considers that these primary pairs account for most of the produced pairs and we neglect secondary photons that can be emitted
by those primary pairs. Note that the number of pairs is not, here, limited to the number $N_0$ of seed electrons as those electrons may emit more than a single photon. This is different from the case of photon-seeded showers [see Fig.~\ref{fig:intro}(b)] for which going beyond the soft-shower limit is needed 
if one wants to enter regimes where the number of produced pairs exceeds that of seed photons. In the photon-seeded model, one thus allows the primary pairs\footnote{As further discussed in Sec.~\ref{sec:model:ps}, we actually consider that only one particle (electron or positron) per primary pair can further contribute to the cascade.} to radiate high-energy photons that can decay into a second generation of pairs. 
Further emission of high-energy photons by this second generation of pairs will however be neglected. We will show in Sec.~\ref{sec:1dpic:Multip_and_plot} that considering only the first two generations provides an excellent approximation in many regimes of interest, and we will give an intuitive justification of why this is so.

Second, we assume that all particles participating in high-energy photon emission and pair production are ultra-relativistic ($\gamma \gg 1$) at all times. 
This allows us to approximate electron and positron velocity as that of light and
to consider that high-energy photons and electron-positron pairs are emitted/created with a momentum aligned with that of the particle they originate from.
We further assume that all particles propagate in the same direction, $-{\bf \hat{z}}$, and will in particular neglect the change of momentum of the charged particles in the laser electromagnetic field. This is a good approximation assuming $\gamma_0 \gg a_0$ (satisfied for parameters of interest for this study).
As a consequence, all particles experience at any given time $t$ the same electric and magnetic fields $\vE(\vz_b(t),t)$ and $\vB(\vz_b(t),t)$,
with $\vz_b(t) = -ct\,{\bf {\hat z}}$ denoting the particles' position at time $t$.
While the model developed in Secs.~\ref{sec:model:es} and~\ref{sec:model:ps} can be applied to arbitrary polarizations, within this work, we will focus on linearly-polarized electromagnetic pulses with an electric field:
\begin{eqnarray}\label{eq:Elaser}
    \vE(z,t) = E_0\,\sin(\omega\,(t-z/c))\,\sin\left(\frac{\pi}{2}\frac{t-z/c}{\tau_p}\right)\,{\bf {\hat y}}\quad{\rm for}\,\, 0\le t-z/c < 2\tau_p\,\,{\rm and}\,\,0\,\,{\rm otherwise}\,,
\end{eqnarray}
and magnetic field $\vB(z,t) = - {\bf \hat{z}} \times \vE/c$. 

Here, we have considered a sin$^2$ intensity profile with $\tau_p$ 
its full-width-at-half-maximum-intensity. 
It follows that knowing a particle's energy at any time $t$
uniquely defines its time-dependent quantum parameter\footnote{Valid for photons as well as electrons or positrons.}
\begin{eqnarray}\label{eq:chi-vs-t}
    \chi(t) = 2\,\frac{\hbar\omega}{m c^2}\,a_0\,\gamma(t)\,\lvert \sin(2\omega t) \rvert \,\sin(\pi t/\tau_p) \quad{\rm for}\,\, 0\le t \le \tau_p\,\,{\rm and}\,\,0\,\,{\rm otherwise}\,.
\end{eqnarray}

Third, the rate of photon emission by electrons and positrons is computed using
the average charged particle energy and the corresponding quantum parameter.
This approximation, which we have found to provide a good prediction for the spectrum of emitted photons, is different from the one considered by Blackburn in 
his model of electron-seeded showers.
Indeed, in Ref.~\cite{blackburn2017scaling}, the authors compute the rate of photon
emission using the initial (seed) electron energy $\gamma_0$ (even though they
do account for energy loss when computing the electron quantum parameter).
This approach can be justified in the low $\chi_0$ regime where 
pair production is dominated by those photons that have been emitted
by the highest energy electrons. It does not hold however in the regime of
interest for his work and radiation losses have to be accounted for.

Last, as we are mostly interested in the range of parameters that will be explored by multi-petawatt laser facilities ($a_0 \gg 1$), 
the processes of high-energy photon emission (nonlinear inverse Compton scattering)
and electron-positron pair production (Breit-Wheeler process)
are described by their (energy) differential rates computed in the locally constant cross-field approximation
(LCFA):
\begin{eqnarray}
    \label{eq:w-nics}w_\chi(\gamma,\gamma_\gamma) &=& \left.\frac{d\Wcs}{d\gamma_\gamma}\right\vert_\chi = \frac{\alpha}{\tau_c}\,\frac{F(\chi,\gamma_\gamma/\gamma)}{\gamma^2}\,,\\
    \label{eq:w-bw}\overline{w}_{\chi_\gamma}(\gamma_\gamma,\gamma) &=& \left.\frac{d\Wbw}{d\gamma}\right\vert_{\chi_\gamma} = \frac{\alpha}{\tau_c}\,\frac{G(\chi_\gamma,\gamma/\gamma_\gamma)}{\gamma_\gamma^2}\,,
\end{eqnarray}
where $w_\chi(\gamma,\gamma_\gamma)\,dt\,d\gamma_\gamma$ denotes the probability for an electron (equivalently positron) of energy $\gamma\,m c^2$ 
and quantum parameter $\chi$ to emit a high-energy photon with energy in between $\gamma_\gamma mc^2$ and $(\gamma_\gamma+d\gamma_\gamma) mc^2$ between times $t$ and $t+dt$,
while $\overline{w}_{\chi_\gamma}(\gamma_\gamma,\gamma)\,dt\,d\gamma$ denotes that of a photon to decay into a pair with the electron carrying an energy 
in between $\gamma mc^2$ and $(\gamma+d\gamma) mc^2$. 
The functions $F(\chi,\xi)$ and $G(\chi_\gamma,\zeta)$ are reported in Appendix~\ref{appendixB}. 
Integrating these differential rates over all possible daughter particle energy 
[photons for Eq.~\eqref{eq:w-nics} and electrons/positrons for Eq.~\eqref{eq:w-bw}] leads to the photon emission and pair production rates
\begin{eqnarray}
    \label{eq:W-nics} \Wcs(\gamma,\chi) = \frac{\alpha}{\tau_c}\,\frac{a(\chi)}{\gamma}\,,\\
    \label{eq:W-bw} \Wbw(\gamma_\gamma,\chi_\gamma) = \frac{\alpha}{\tau_c}\,\frac{b(\chi_\gamma)}{\gamma_\gamma}\,,
\end{eqnarray}
where the functions $a(\chi)$ and $b(\chi_\gamma)$ and their asymptotic limits are reported in Appendix~\ref{appendixB}.
In Fig.\ref{fig:intro}(c), black and red dotted lines report the values of ($a_0$,$\gamma_0$) for which $\Wcs(\gamma_0,\chi_0)=\omega$ and $\Wbw(\gamma_0,\chi_0)=\omega$, respectively. The values of ($a_0$,$\gamma_0$) for which $\Wbw(\gamma_0,\chi_0)=1/\tau_p$ 
(with $\tau_p=20~{\rm fs}$) are also reported as a blue solid line defining the regime of validity of the model proposed by Blackburn et al.~\cite{blackburn2017scaling}.
It is clear that, in the regime of interest for future experiments 
at multi-petawatt laser facilities, abundant high-energy photon emission 
is expected on timescales of the order of or shorter than the optical cycle ($\Wcs > \omega$). In contrast, the characteristic time for
a photon to decay into an electron-positron pair will be larger than the 
optical cycle, yet shorter than the characteristic pulse duration (considering $\tau_p \ge 20~{\rm fs}$).\\

In what follows, we describe the models that have been developed for electron-seeded and photon-seeded cascades. In both cases, we first give the governing equations for the distribution functions of the successive generation of species (electrons, positrons and photons) participating in the shower. We then derive our model and discuss our assumptions. Last we briefly summarise how the model is used to compute the final number of pairs produced per incident (seed) particle, henceforth referred to as the shower multiplicity.

\subsection{Electron seeding}\label{sec:model:es}

When the shower is seeded by an incident electron beam, we need to follow the dynamics of the seeding electrons through their distribution function
$f^{(0)}(\gamma,t)$, the first generation of gamma photons, 
i.e. those emitted by the seeding electron, 
though their distribution function $f_\gamma^{(1)}(\gamma_\gamma,t)$,
and the first generation of created pairs through their distribution function $f_\pm^{(1)}(\gamma,t)$.
The equations of evolution of these distribution functions 
read\footnote{As stressed in Sec.~\ref{sec:model:hyp}, we neglect
the work of the laser electric field on all charged particles.}:
\begin{eqnarray}
    \nonumber \partial_t f^{(0)}(\gamma,t) &=& \int_0^{\infty}\!\! d\gamma_\gamma\, w_\chi(\gamma+\gamma_\gamma,\gamma_\gamma)\,f^{(0)}(\gamma+\gamma_\gamma,t)\\
    \label{eq:es:f0}&-& \int_0^{\infty}\!\! d\gamma_\gamma\, w_\chi(\gamma,\gamma_\gamma)\,f^{(0)}(\gamma,t)\,,\\
    \nonumber \partial_t f_\gamma^{(1)}(\gamma_\gamma,t) &=& \int_1^{\infty}\!\! d\gamma\, w_\chi(\gamma,\gamma_\gamma)\,f^{(0)}(\gamma,t)\\
    \label{eq:es:fg1}&-& \int_1^{\infty}\!\! d\gamma\, \overline{w}_{\chi_\gamma}(\gamma_\gamma,\gamma)\,f_\gamma^{(1)}(\gamma_\gamma,t)\,,\\
    \nonumber \partial_t f_\pm^{(1)}(\gamma,t) &=& \int_0^{\infty}\!\! d\gamma_\gamma\, \overline{w}_{\chi_\gamma}(\gamma_\gamma,\gamma)\,f_\gamma^{(1)}(\gamma_\gamma,t)\\
    \nonumber &+& \int_0^{\infty}\!\! d\gamma_\gamma\, w_\chi(\gamma+\gamma_\gamma,\gamma_\gamma)\,f_\pm^{(1)}(\gamma+\gamma_\gamma,t)\\
    \label{eq:es:fpm1}&-& \int_0^{\infty}\!\! d\gamma_\gamma\, w_\chi(\gamma,\gamma_\gamma)\,f_\pm^{(1)}(\gamma,t)\,.
\end{eqnarray}
Note that all differential rates are time-dependent through their dependency in $\chi$.

Equation \eqref{eq:es:f0} describes the cooling of the seeding electron beam as it collides 
with the laser pulse and electrons emit high-energy gamma photons (see, e.g., Ref.~\cite{niel2021classical}). 
As will be demonstrated in Sec.~\ref{sec:1dpic:Multip_and_plot}, in the parameter range of interest for this study,
it is not mandatory to capture the details of the seed electron distribution function $f^{(0)}$.
It is sufficient to compute the time-dependent seed electron average energy:
\begin{eqnarray}\nonumber
    \langle\gamma\rangle_{_{\!0}}(t) \equiv N_0^{-1}\,\int_1^\infty\!\!d\gamma\,\gamma\,f^{(0)}(\gamma,t)\,,
\end{eqnarray}
which temporal evolution is well approximated by ~\cite{niel2021classical}:
\begin{eqnarray}\label{eq:es:gamma_moy_vs_time}
    \frac{d}{dt}\langle\gamma\rangle_{_{\!0}} \simeq -W_0\,\langle\chi\rangle_{_{\!0}}^2 \, g\!\left(\langle\chi\rangle_{_{\!0}}\right),
\end{eqnarray}
with $W_0=2\alpha/(3\tau_c)$ and $\langle\gamma\rangle_{_{\!0}}(t=0)=\gamma_0$. 
Here $\langle\chi\rangle_{_{\!0}}(t)$ denotes the quantum parameter\footnote{Note that here $\langle\chi\rangle_{_{\!0}}(t=0) = 0 \neq \chi_0$ with $\chi_0$ defined by Eq.~\eqref{eq:chi0}.} of a charged particle (either electron or positron)
with energy $\langle\gamma\rangle_{_{\!0}}mc^2$ computed at time $t$ [using Eq.~\eqref{eq:chi-vs-t}], and
\begin{eqnarray}\nonumber
    g(\chi) = \frac{2}{3\chi^2}\,\int_0^1\!d\xi\,\xi\,F(\chi,\xi)
\end{eqnarray}
is the quantum correction (Gaunt factor) on the power radiated away by an ultra-relativistic electron 
in a strong field\footnote{In our model, we use the approximate form proposed by Baier et al.~\cite{baier1998electromagnetic}:
$g(\chi) \simeq [1+4.8(1+\chi)\ln(1+1.7\chi)+2.44\chi^2]^{-2/3}$.}. 

Let us note that Eq.~\eqref{eq:es:gamma_moy_vs_time} is the same as Eq.~(6) in~\cite{blackburn2017scaling}. 
It was solved analytically by Blackburn et al. in the limit of small quantum parameters and pair production probability, and considering a laser pulse with a Gaussian temporal profile.
This solution is however not valid in the regime of interest for this work and Eq.~\eqref{eq:es:gamma_moy_vs_time} will be solved numerically.

Equation~\eqref{eq:es:fg1} describes the evolution of the first generation of high-energy photons. Those photons are emitted by the seed electrons as described by the first term in the right-hand-side (rhs) and may decay into electron-positron pairs as described by the second term. 
We now rewrite this equation 
introducing the pair production rate $\Wbw(\gamma_\gamma,\chi_\gamma)$ [Eq.~\eqref{eq:W-bw}] 
and considering that the emission of high-energy photons is dominated by seed electrons at the average energy $\langle\gamma\rangle_{_{\!0}}$,
i.e. using $f^{(0)} \simeq N_0\,\delta\left(\gamma - \langle\gamma\rangle_{_{\!0}}(t)\right)$ in the first term:
\begin{eqnarray}\nonumber
    \partial_t f_\gamma^{(1)} = N_0\,w_\chi\!\left(\langle\gamma\rangle_{_{\!0}}(t),\gamma_\gamma\right)-\,\Wbw(\gamma_\gamma,\chi_\gamma(t))\,f_\gamma^{(1)}\,,
\end{eqnarray}
with $\chi_\gamma(t)$ the quantum parameter of a photon with energy $\gamma_\gamma$ computed at time $t$ using Eq.~\eqref{eq:chi-vs-t}.
This leads to:
\begin{eqnarray}\label{eq:es:fg1-solution}
    f_\gamma^{(1)}(\gamma_\gamma,t) = N_0\,A(\gamma_\gamma,t)\,\exp\left(-\int_0^t\!d\tau\,\Wbw(\gamma_\gamma,\chi_\gamma(\tau))\right)\,
\end{eqnarray}
with
\begin{eqnarray}\nonumber
    A(\gamma_\gamma,t) = \int_0^t\!dt'\,w_\chi\!\left(\langle\gamma\rangle_{_{\!0}}(t'),\gamma_\gamma\right)\,
    \exp\bigg(\int_0^{t'}\!d\tau\,\Wbw(\gamma_\gamma,\chi_\gamma(\tau))\bigg) \,.
\end{eqnarray}

Finally, Eq.~\eqref{eq:es:fpm1} describes the evolution of the first generation of pairs produced from 
the decay of the (first generation) high-energy photons as described by the first term in the equation's rhs.
The last two terms account for the modification of the pair energy spectrum due to high-energy photon
emission (radiation reaction). The temporal evolution of the number of pairs $N_{\pm}^{(1)}$ of the first generation 
is computed integrating\footnote{Note that the last two terms in Eq.~\eqref{eq:es:fpm1} cancel 
upon integration in $\gamma$ as high-energy photon emission conserves the number of charges.} over all $\gamma$,
which, after time integration leads to:
\begin{eqnarray}\nonumber
    N_\pm^{(1)}(t) = \int_0^\infty\!d\gamma_\gamma\,\int_0^t\!dt'\,\Wbw(\gamma_\gamma,\chi_\gamma(t'))\,f_\gamma^{(1)}(\gamma_\gamma,t')\,.
\end{eqnarray}
Using Eq.~\eqref{eq:es:fg1-solution} for $f_\gamma^{(1)}$ and integrating by part finally leads:
\begin{eqnarray}\label{eq:es:nbofpairs}
    N_\pm^{(1)}(t) = N_0\,\int_0^\infty\!d\gamma_\gamma\,
    \int_0^t\!dt'\,w_\chi(\langle\gamma\rangle_{_{\!0}}(t'),\gamma_\gamma)\,
    \left[1-\exp\left(-\int_{t'}^t\!d\tau\,\Wbw(\gamma_\gamma,\chi_\gamma(\tau))\right)\right]\,.
\end{eqnarray}

The final number of produced pairs, approximated as $N_\pm^{(1)}$, 
can then be obtained integrating Eq.~\eqref{eq:es:nbofpairs} numerically.

\subsection{Photon seeding}\label{sec:model:ps}

When the shower is seeded by a high-energy photon beam, we need to follow the dynamics
of this photon beam through its distribution function $f^{(0)}(\gamma_\gamma,t)$, 
and at least the first generation of pairs it can produce.
However, rather than following the full pair distribution function $f_\pm^{(1)}(\gamma,t)$
[as described by Eq.\eqref{eq:es:fpm1}],
we focus our attention only on those particles, either electrons or positrons,
that carry the most energy at their moment of creation.
Indeed, simulations presented in Sec.~\ref{sec:pic1d} show that those particles,
hereinafter described by their distribution function $\hat{f}_\pm^{(1)}(\gamma,t)$, 
are the ones that can radiate secondary photons, described by $f_\gamma^{(1)}(\gamma_\gamma,t)$,
that in turn will create more electron-positron pairs, described by their distribution function $f_\pm^{(2)}(\gamma,t)$. 
The equations of evolution of these distribution functions read:
\begin{eqnarray}
    \label{eq:ps:f0}\partial_t f^{(0)}(\gamma_\gamma,t) &=& -\int_1^\infty\!d\gamma\,\overline{w}_{\chi_\gamma}(\gamma_\gamma,\gamma)\,f^{(0)}(\gamma_\gamma,t)\,,\\
    \nonumber\partial_t \hat{f}_\pm^{(1)}(\gamma,t) &=& 2\int_0^\infty\!d\gamma_\gamma\,\overline{w}_{\chi_\gamma}(\gamma_\gamma,\gamma)\,H(\gamma-\gamma_\gamma/2)\,f^{(0)}(\gamma_\gamma,t)\\
    \nonumber&+& \int_0^\infty\!d\gamma_\gamma\,w_{\chi}(\gamma+\gamma_\gamma,\gamma_\gamma)\,\hat{f}_\pm^{(1)}(\gamma+\gamma_\gamma,t)\\
    \label{eq:ps:fpm1}&-& \int_0^\infty\!d\gamma_\gamma\,w_{\chi}(\gamma,\gamma_\gamma)\,\hat{f}_\pm^{(1)}(\gamma,t)\,,\\
    \nonumber\partial_t f_\gamma^{(1)}(\gamma_\gamma,t) &=& \int_1^\infty\!d\gamma\,w_{\chi}(\gamma,\gamma_\gamma)\,\hat{f}_\pm^{(1)}(\gamma,t)\\
    \label{eq:ps:fg1}&-&\int_1^\infty\!d\gamma\,\overline{w}_{\chi_\gamma}(\gamma_\gamma,\gamma)\,f_\gamma^{(1)}(\gamma_\gamma,t)\,,\\
    \nonumber\partial_t f_\pm^{(2)}(\gamma,t) &=& \int_0^\infty\!d\gamma_\gamma\,\overline{w}_{\chi_\gamma}(\gamma_\gamma,\gamma)\,f_\gamma^{(1)}(\gamma_\gamma,t)\\
    \nonumber&+& \int_0^\infty\!d\gamma_\gamma\,w_{\chi}(\gamma+\gamma_\gamma,\gamma_\gamma)\,f_\pm^{(2)}(\gamma+\gamma_\gamma,t)\\
    \label{eq:ps:fpm2}&-& \int_0^\infty\!d\gamma_\gamma\,w_{\chi}(\gamma,\gamma_\gamma)\,f_\pm^{(2)}(\gamma,t)\,,
\end{eqnarray}
where $H(x)$ denotes the Heaviside function. 

Equation~\eqref{eq:ps:f0} describes the decay of the incident photons into electron-positron pairs as they collide with the laser pulse.
Considering an initially mono-energectic photon beam, the solution of Eq.~\eqref{eq:ps:f0} can be written in the form:
\begin{eqnarray}\nonumber
    f^{(0)}(\gamma_\gamma,t) = N_0\,\exp\left(-\int_0^t\!d\tau\,\Wbw(\gamma_0,\chi_{\gamma_0}(\tau))\right)\,\delta(\gamma_\gamma-\gamma_0)\,,
\end{eqnarray}
where $\chi_{\gamma_0}(t)$ is the quantum parameter of a seed photon with energy $\gamma_0 mc^2$, computed at time $t$ using Eq.~\eqref{eq:chi-vs-t}. 

Equation~\eqref{eq:ps:fpm1} describes the evolution of the resulting (first) generation of pairs as they are created (first term in the equation's rhs)
and then emit high-energy radiation (the last two terms in the equation's rhs). As stressed earlier, only those particles (either electrons or positrons)
that are created with the most energy are accounted for here. 
Simulations\footnote{Let us add that considering the full distribution function of primary pairs $f_\pm^{(1)}$ (rather than $\hat{f}_\pm^{(1)}$) and then considering that pairs at the average energy dominate the emission of (secondary) high-energy photons does not lead a good agreement with PIC simulations.} indeed show that, of the primary pair particles, those (either electron or positron) which have been created with the highest energy contribute most to the following development of the shower. 
This can be understood as follows. 
In the small-$\chi$ limit, both electron and positron are statistically created with a similar yet different energy, equivalently quantum parameter.
Because of exponential suppression of the pair production rate in the small-$\chi$ limit, the particle with the highest energy/quantum parameter is strongly advantaged compared to the other one in further participating in the shower. 
In the large-$\chi$ limit, the pair production rate is such that, at the moment the electron-positron pair is created, most of the decayed photon energy is carried away by one or the other particle. 

Consistently with Eq.~\eqref{eq:ps:f0}, Eq.~\eqref{eq:ps:fpm1} provides the time evolution of
the number of pairs of the first generation:
\begin{eqnarray}\label{eq:ps:Npm1}
    N_\pm^{(1)}(t) = N_0\,\left[1 - \exp\left(-\int_0^t\!d\tau\,\Wbw(\gamma_0,\chi_{\gamma_0}(\tau))\right)\right]\,.
\end{eqnarray}
An analytical expression of the integral has been proposed in \cite{mercuri2021impact} which is reproduced in Appendix~\ref{appendixC}  for the case of Eq.~\eqref{eq:Elaser}. In what follows we will however integrate this equation numerically.

Proceeding as in Sec.~\ref{sec:model:es}, we will focus our attention on those particle mean energy.
Multiplying Eq.~\eqref{eq:ps:fpm1} by $\gamma$ and integrating over all $\gamma$ provides us with the equation
for the average energy $\langle\gamma\rangle_{\!_1}(t)$ for $t>0$\footnote{Formally, the term $\left[\gamma_0 - \langle\gamma\rangle_{\!_1}(t)\right]$ in the rhs of Eq.~\eqref{eq:ps:gamma1} should have been written $\left[\gamma_{\rm cr}(t) - \langle\gamma\rangle_{\!_1}(t)\right]$ with $\gamma_{\rm cr}(t) = \int_{\gamma_0/2}^{\gamma_0}\!d\gamma\,\gamma\,\overline{w}(\gamma_\gamma,\gamma)\,\big/\int_{\gamma_0/2}^{\gamma_0}\!d\gamma\,\overline{w}(\gamma_\gamma,\gamma)$. Extensive tests have shown that, while using $\gamma_{\rm cr}(t)$  provides fair predictions 
for the final number of pairs, a better agreement is found using $\gamma_{\rm cr}(t) = \gamma_0$ at all times.}:
\begin{eqnarray}\label{eq:ps:gamma1}
    \frac{d}{dt}\langle\gamma\rangle_{\!_1} = -W_0\,\langle\chi\rangle_{1}^2(t)\,g\!\left(\langle\chi\rangle_{1}(t)\right) 
    + \Wbw(\gamma_0,\chi_{\gamma_0}(t))\Big[ \gamma_0 - \langle\gamma\rangle_{\!_1}(t) \Big]\,
    \frac{N_0-N_\pm^{(1)}(t)}{N_\pm^{(1)}(t)}
\end{eqnarray}
with $\langle\chi\rangle_{1}(t)$ the quantum parameter of an electron with the energy $\langle\gamma\rangle_{\!_1}(t)$
computed at time $t$ using Eq.~\eqref{eq:chi-vs-t} and\footnote{At the very beginning of the interaction ($t \sim 0$), electrons experience
 a very small electromagnetic field, hence $\chi_{\gamma_0} \rightarrow 0$. In the unlikely case a pair is created, it is thus very probable that the photon energy will be equally split between the electron and positron.}
 $\langle\gamma\rangle_{\!_1}(t=0)=\gamma_0/2$.

Equation~\eqref{eq:ps:fg1} then describes the evolution of the energy distribution of secondary photons that, on the one hand, 
are created by those (highest energy) first generation of electron-positron pairs [first term in the rhs of Eq.~\eqref{eq:ps:fg1}]
and, on the other hand, decay into secondary electron-positron pairs [second term in the rhs of Eq.~\eqref{eq:ps:fg1}].
This equation can be solved considering that photon emission is dominated by electrons and positrons at the average mean energy, i.e. using $\hat{f}_\pm^{(1)}(\gamma,t) \simeq N_\pm^{(1)}(t)\,\delta\left(\gamma-\langle\gamma\rangle_{\!_1}(t)\right)$, 
leading to:
\begin{eqnarray}\label{eq:ps:fg1-sol}
    f_\gamma^{(1)}(\gamma_\gamma,t) = B(\gamma_\gamma,t)\,\exp\left(-\int_0^t\!d\tau\,\Wbw(\gamma_\gamma,\chi_\gamma(\tau))\right)
\end{eqnarray}
with
\begin{eqnarray}\nonumber
    B(\gamma_\gamma,t) = \int_0^t\!dt'\,N_\pm^{(1)}(t')\,w_\chi(\langle\gamma\rangle_{\!_1}(t'),\gamma_\gamma)\,
    \exp\left(\int_0^{t'}\!d\tau\,\Wbw(\gamma_\gamma,\chi_\gamma(\tau))\right)\,.
\end{eqnarray}

Finally, Eq.~\eqref{eq:ps:fpm2} describes the evolution of the secondary pairs.
Using Eq.~\eqref{eq:ps:fg1-sol}, it gives, after integration over $\gamma$, the numbers of secondary pairs:
\begin{eqnarray}\label{eq:ps:Npm2}
    N_\pm^{(2)}(t) = \int_0^\infty\!d\gamma_\gamma\int_0^t\!dt'\,N_\pm^{(1)}(t')\,w_\chi\!\left(\langle\gamma\rangle_{\!_1}(t'),\gamma_\gamma\right)
    \,\left[1 - \exp\left(-\int_{t'}^t\!d\tau\,\Wbw(\gamma_\gamma,\chi_\gamma(\tau))\right)\right]\,.
\end{eqnarray}\\

The total number of pairs is finally computed summing the contribution of primary and secondary pairs.
While an analytical approximation for the number of primary pairs $N_\pm^{(1)}$ was proposed in Ref.~\cite{mercuri2021impact} (see also Appendix~\ref{appendixC}, Eq. (2C)),
we here integrate Eq.~\eqref{eq:ps:Npm1} numerically. 
We proceed similarly with Eq.~\eqref{eq:ps:gamma1} (for $\langle\gamma\rangle_{\!_1}$) and
Eq.~\eqref{eq:ps:Npm2}  (for $N_\pm^{(2)}$).

\section{One-dimensional PIC simulations}\label{sec:pic1d}

\subsection{Simulation parameters}

In this Section, we discuss the results of a series of one-dimensional (written 1D3V since velocities are three-dimensional) PIC simulations that follow the head-on collision 
of a thin (two-cell wide) monoenergetic beam of seed particles (either electrons or photons)
against an ultra-intense laser pulse. 
Seed particles are initialized as a mono-energetic beam, all particles of the beam propagating in the same direction and 
having the same initial energy $\mathcal{E}_0  = (\gamma_0-1) m c^2$ for electrons and $\mathcal{E}_0  = \gamma_0 m c^2$ for photons. 
The intense laser pulse is a linearly polarized plane wave with maximum relativistic field strength parameter $a_0 = e E_0/(m c \omega)$ ($E_0$ the maximum electric field),
and a sin$^2$ intensity profile with full-width-at-half-maximum duration $\tau_p$, as described by Eq.~\eqref{eq:Elaser}. 
In general, and unless specified otherwise, two types of laser systems are discussed: 
ultra-short (20 fs) Ti:Sapphire ($\lambda=0.8 {\rm \mu m}$) pulses, 
and longer (150 fs) Nd-glass ($\lambda=1.057 {\rm \mu m}$) pulses. 

All simulations were performed with the open-source code Smilei~\cite{smilei_derouillat_2018} that embarks Monte-Carlo modules\footnote{For better 
accuracy, precomputed tables with 1024 points were used as described in \url{https://smileipic.github.io/Smilei/Use/tables.html}.}
to account for high-energy photon emission (nonlinear inverse Compton scattering) and pair production (nonlinear Breit-Wheeler).
For those simulations, the spatial and temporal resolution were $\Delta x = ~\lambda/64$ and $c\Delta t = ~\Delta x/2$, 
respectively. All $2\times10^4$ (if $\chi_0>2$) or $2\times10^6$ (if $\chi_0\leq2$) seed macro-particles were initialized in two cells.
The simulations were run long enough that no more particles (either photons or pairs) were created by the end of each simulation.

\subsection{Multiplicity as a function of $a_0$ and $\gamma_0$  and comparison to the reduced model}\label{sec:1dpic:Multip_and_plot}

\begin{figure*}[t]
    \includegraphics[width=1.\linewidth]{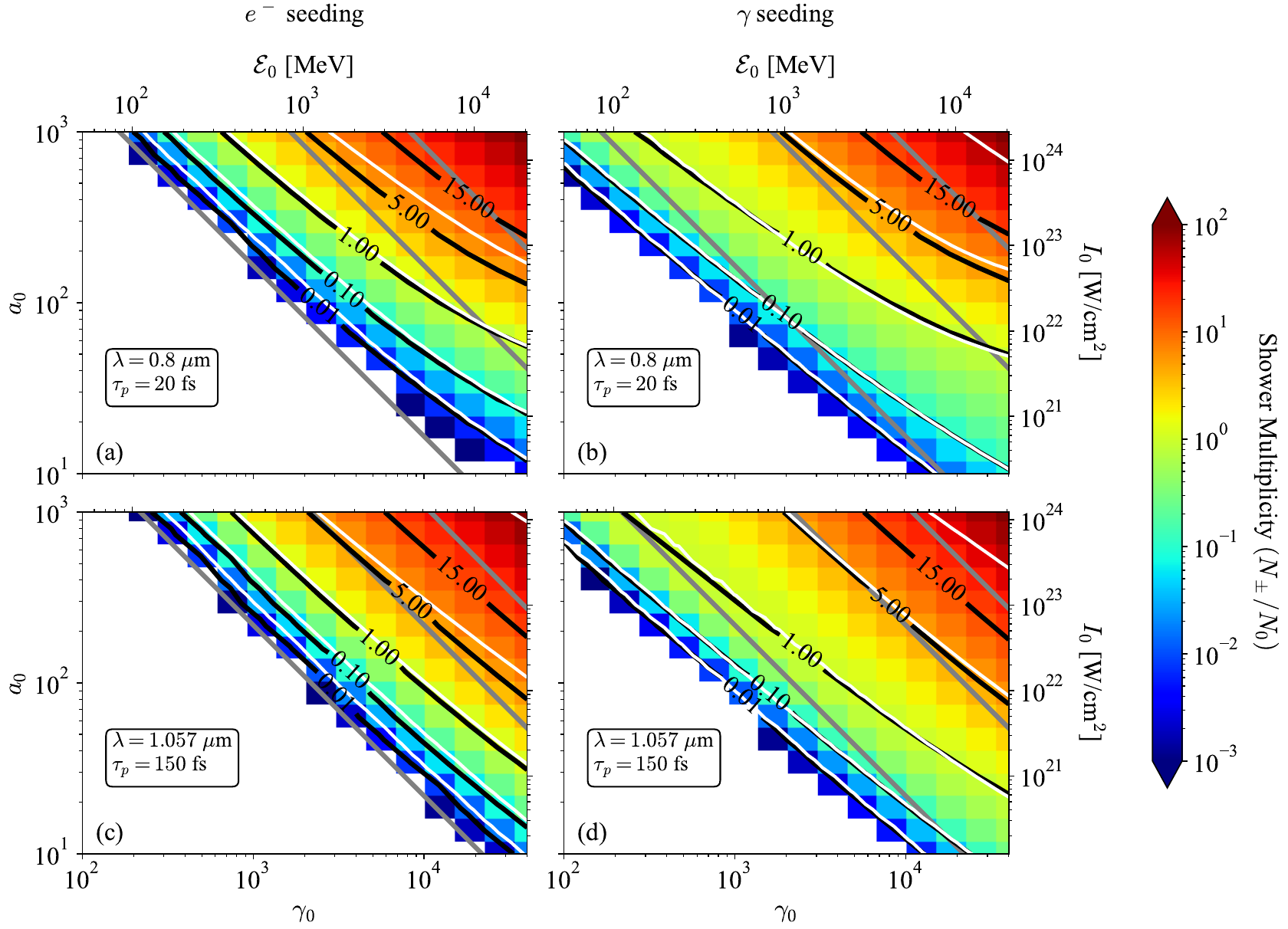}
        \caption{Shower multiplicity, i.e. final number of pairs per incident particle, extracted from 1D3V PIC simulations as a function of the laser field strength $a_0$ (equivalently intensity $I_0$) 
        and incident particle normalized energy $\gamma_0$ (equivalently $\mathcal{E}_0$). Considering electron seeding (left panels) 
        and photon seeding (right panels). 
        For short pulse ($\tau_p=20$ fs) Ti:Sapphire lasers (top panels) and longer pulse ($\tau_p=150$ fs) Nd:glass (bottom panels). 
        Black solid lines report iso-contours extracted from simulations. 
        White solid lines report predictions from this work [Eq. \eqref{eq:es:nbofpairs} for electron seeding 
        and Eqs. \eqref{eq:ps:Npm1} and \eqref{eq:ps:Npm2} for photon seeding]. 
        Grey lines stand for $\chi_0 = 1$, $10$ and $50$ (from bottom-left to top-right).} 
        \label{fig:1d3vSim:NbOfPairs}
    \end{figure*}

In Fig.~\ref{fig:1d3vSim:NbOfPairs}, we report, as a function of $\gamma_0$ and $a_0$, the multiplicity ($N_\pm/N_0$)
extracted from a set of 1D3V PIC simulations of electron-seeded (left panels) and photon-seeded (right panels) showers,
for ultra-short ($20~{\rm fs}$, $\lambda = 0.8~{\rm \mu m}$) pulses (top panels) and longer ($150~{\rm fs}$, $\lambda = 1.057~{\rm \mu m}$) pulses (bottom panels).
Each panel summarizes the results of $256$ simulations considering $16$ values of $a_0$ from $10$ to $1000$
and $16$ values of $\gamma_0$ from $100$ to $4\times 10^4$ (both axes discretized in logarithmic scale).

The white region in the bottom-left part of each panel corresponds to a low multiplicity regime ($N_\pm/N_0 \lesssim {10^{-3}}$),
that will not be much discussed here for two reasons. 
First, it is of marginal interest for forthcoming experiments on multi-petawatt laser systems, see Fig.~\ref{fig:intro}.
Second, it corresponds to a regime where the standard Monte-Carlo approach implemented in Smilei is not optimal 
(at least for systematic studies performed without consuming too much computer resources).
Indeed, Monte-Carlo methods for the description of photon disintegration require a particle number greater than $N$ 
to ensure precision at $1/\sqrt{N}$ level. As our simulations were performed with $2\times10^6$ initial particles, 
the low multiplicity regime ($\lesssim 10^{-3}$) is not well described.

The color-scaled region thus reports the number of produced pairs per incident particle, or multiplicity, 
as small as {$10^{-3}$} at moderate $\chi_0 \lesssim 1$, and up to near a hundred at very large $\chi_0$ (top-right corner).
Iso-contours of multiplicity $N_\pm/N_0 = 10^{-2}$, $10^{-1}$, $1$, $5$ and $15$ are also reported: 
in black when extracted from our PIC simulations, and in white for the model predictions.
For electron-seeded showers, a very good agreement is observed up to multiplicities of $N_\pm/N_0 = 2$ (not shown) and fair agreement for multiplicities of $5$.
For photon-seeded showers, a very good agreement is observed up to multiplicities of $N_\pm/N_0 = 5$. 

Before discussing the differences between electron- and photon seeding, as well as ultra-short (20 fs) and longer (150 fs) pulses, 
let us note that the multiplicity iso-contours are not, in general, straight lines parallel to iso-$\chi_0$ (gray dotted) contours.
This means that $\chi_0$ is not the only parameter determining the total number of pairs, 
and that by extension $\gamma_0$ and $a_0$ do not play a symmetrical role in the shower development.
This was already discussed for the case of photon-seeded showers in \cite{mercuri2021impact},
and could have been expected from Eqs.~\eqref{eq:W-nics} and~\eqref{eq:W-bw} 
where the rates depend on the incident particle energy as well as on its quantum parameter.\\ 

\begin{figure*}
\includegraphics[width=1.\linewidth]{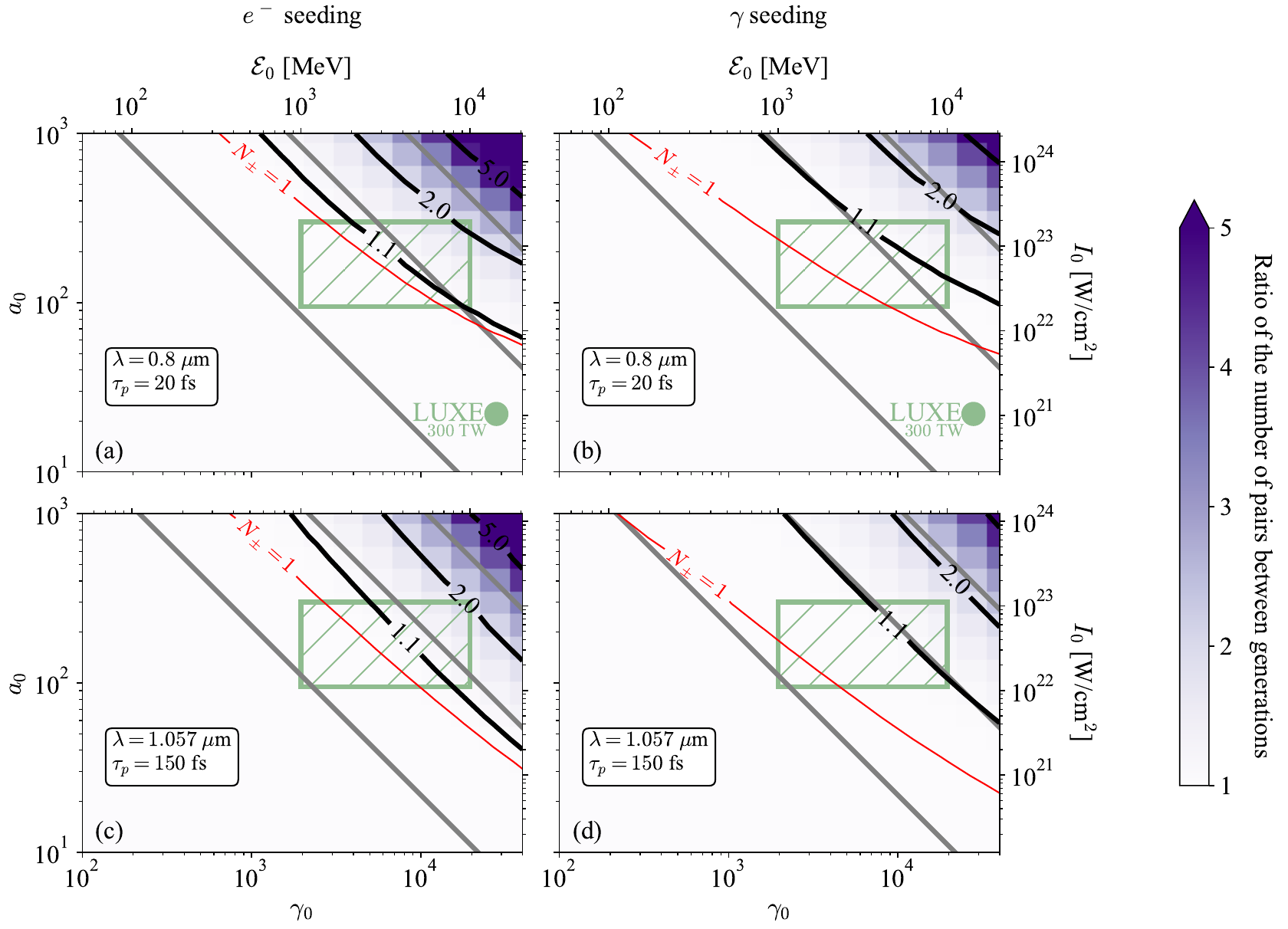}
    \caption{Ratio between the total number of pairs (accounting for all generations) and the number of pairs produced during the first (only, for electron seeding) and second (for photon seeding) generations as a function of the laser field strength $a_0$ and incident particle normalized energy $\gamma_0$. For short pulse ($\tau_p=20$ fs) Ti:Sapphire lasers (top panels) and long pulse ($\tau_p=150$ fs) Nd:Glass (bottom panels). Black solid lines report iso-contours extracted from simulations. Red dashed lines stand for $N_\pm=1$. The green hatched region corresponds to the region of interest for PW to multi-PW laser systems. Also reported on top panels as a green dot is an operating regime for the LUXE $300$ TW experiment. Grey lines stand for $\chi_0=1$, $10$ and $50$ form (from bottom left to top right).} \label{fig:number_species_smilei}
\end{figure*}

We are now in a position to discuss the model assumptions, namely the fact that we limit ourselves to a few generations and that to describe the energy evolution of the leptons we don't need the full distribution function but we can limit ourselves to the first moment i.e. the average energy. In Fig.~\ref{fig:number_species_smilei} we present (with the same hierarchy defined in Fig.~\ref{fig:1d3vSim:NbOfPairs}) the final number of pairs divided by the number of pairs produced from the first (only for electron seeding) and the second (for photon seeding) generations extracted from PIC simulations. This plot acts as a definition of the regime of validity of our model where the black solid line (or the color map) corresponds to the iso-contour equal to $1.1,2,5$ and represents the error factor induced by the consideration of the first generations only. It is clear that for future experimental parameters, only a few generations of creation: first for electron seeding and first and second for photon seeding are enough to describe the number of pairs with at most an error of a factor $2$.

Even if $\chi$ is not the only relevant parameter and because our results have a weak dependence on the pulse duration, at zero order, pair creation will not happen below a threshold of $\chi=1$. Moreover as discussed in Fig.~\ref{fig:number_species_smilei} our model is very well satisfied for $\chi_0$ less than $\approx 20$. We can thus consider a simple argument to estimate the number of relevant generations \cite{akhiezer1994kinetic}. If $\chi\gg1$ on average an initial electron with an initial energy of $\gamma_0$ emits a photon with an energy of $\approx \gamma_0/4$. This assumption holds for a large range of $\chi_0$ and is still approximately correct down to $\chi=1$ (the average energy varying from $\approx \gamma_0/4$ at $\chi\gg 1$ to $\approx \gamma_0/7$ at $\chi=1$). The produced photon will create the first pair so that all its energy goes to one particle that will again emit a photon of energy $\approx \gamma_0/16$. The reaction chain stopping for $\chi=1$, we can infer that the initial value of $\chi_0$ consistent with one generation will be of the order of $16$. For an initial photon at $\gamma_0$, the reasoning is similar. This photon decay to create one electron with the same energy $\gamma_0$ and no one to the positron. Using the same intuitive result, the third generation starts for $\chi_0$ larger or order of $16$. Even if all the parameters are not considered here, this simple estimation is in agreement with the full numerical analysis presented in Fig.~\ref{fig:number_species_smilei} and allows us to get a rough estimate of the range of validity of the model.

The comparison between the model and PIC simulations shown in Fig.~\ref{fig:1d3vSim:NbOfPairs} proves that the average energy approximation gives a good estimation of the number of pairs. The initial particles are mono-energetic so at the beginning of the interaction the average energy is enough to describe the photon emission. After a certain cooling characteristic time, the spectrum also tends to a sharp Gaussian confirming the average approximation. During the transition regime, the spectrum is particularly broad but the average approximation still well describes the lower energetic photons emission. By using this hypothesis, we always underestimate the number of emitted photons with the energy of the order of their parent electron. This is a problem only in low multiplicity regimes ($\lessapprox 0.1$) where the only photons capable of decay are those that have an energy near to their parent electron. Even if the model well predicts the scaling of the number we get an underestimation at low multiplicity. While in multiplicity order of $1$, these rare photons do not influence the number of pairs because the decay process is mainly due to a high number of photons with lower energy. The average energy approximation allows first to take into account the energy conservation in the photon emission and second to describe these low energetic photons with a good precision giving an estimation in excellent agreement with simulations for moderate multiplicity (order of $1$) regime.

\subsection{electron seeding vs photon seeding}

\begin{figure*}
    \includegraphics[width=1\linewidth]{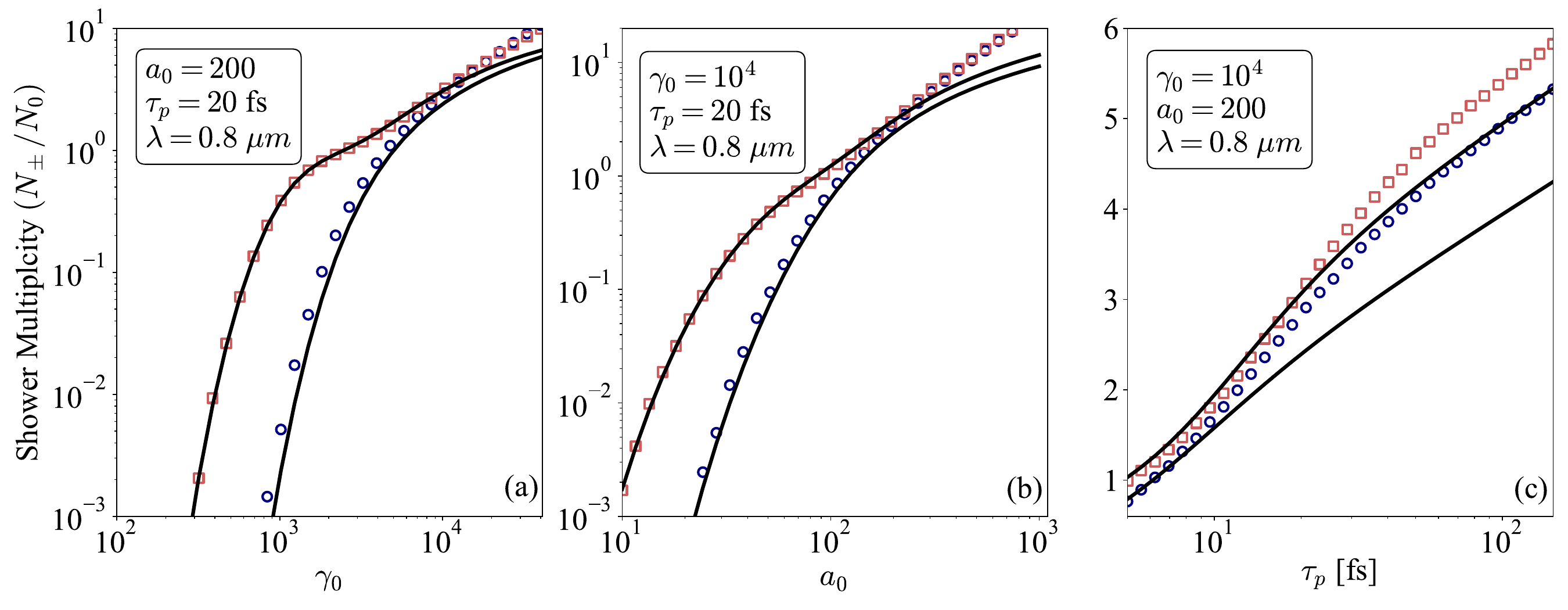}
    \caption{Shower multiplicities extracted from 1D3V PIC simulations for electron seeding (red squares) and photon seeding (blue circles) accounting for all generations. Black solid lines represent the predictions from our model [Eq. \eqref{eq:es:nbofpairs} for electron seeding and Eqs. \eqref{eq:ps:Npm1} and \eqref{eq:ps:Npm2} for photon seeding]. Shown as a function of the incident particle energy $\gamma_0$ (a), the laser field strength $a_0$ (b), and the pulse duration (c).}
    \label{fig:dependences_1D}
\end{figure*}

Over the whole range of parameters studied here,  we find that photon seeding always 
provides multiplicities larger or equal to those obtained using electron seeding. However, forthcoming experiments will likely produce a significantly larger number of GeV-level electrons than photons of similar energy. Apart from this practical consideration, it is interesting to explain the difference between electron and photon seeding. For clarity, we plot in Fig.~\ref{fig:dependences_1D}(a) the multiplicity of pairs as a function of $\gamma_0$ for $a_0=200$, $\tau_p=20$fs and $\lambda=0.8 \mu$m  in the case of electron seeding (blue circles) and photon seeding (red squares). The black lines report our theoretical predictions obtained from Eq.~\eqref{eq:es:nbofpairs} (for electron seeding) and Eqs.~\eqref{eq:ps:Npm1} and~\eqref{eq:ps:Npm2} (for photon seeding). The same quantity is shown as a function of $a_0$ for $\gamma_0=10^4$, $\tau_p=20$fs and $\lambda=0.8 \mu$m in panel~(b).

The largest difference between the two seeding strategies appears in the low multiplicity regime [$\gamma_0<10^3$ in Fig.~\ref{fig:dependences_1D}(a) or $a_0<100$ in Fig.~\ref{fig:dependences_1D}(b)] where the number of produced pairs per incident electron is significantly smaller than 
that obtained seeding the shower with photons. This can be easily explained as, in this regime, the seed electrons 
will emit many photons but with an energy significantly smaller than $\gamma_0$. Their quantum parameter will be much less than the quantum parameter $\chi_0$ associated to $\gamma_0$. The probability of these 
low-energy photons to decay will thus be significantly (exponentially) suppressed compared to the initial photon seeding case.

In contrast, in the large multiplicity regime, there is virtually no difference between seeding the shower with an electron or a photon beam at the same initial energy. A seed photon will quickly turn into a pair with one of the particles (either the electron or the positron)
carrying most of its energy, so the situation is very similar to seeding the shower with an electron beam.\\

\subsection{Ultra-short vs longer pulses} 
\begin{figure}
    \centering
    \includegraphics[width=1.\linewidth]{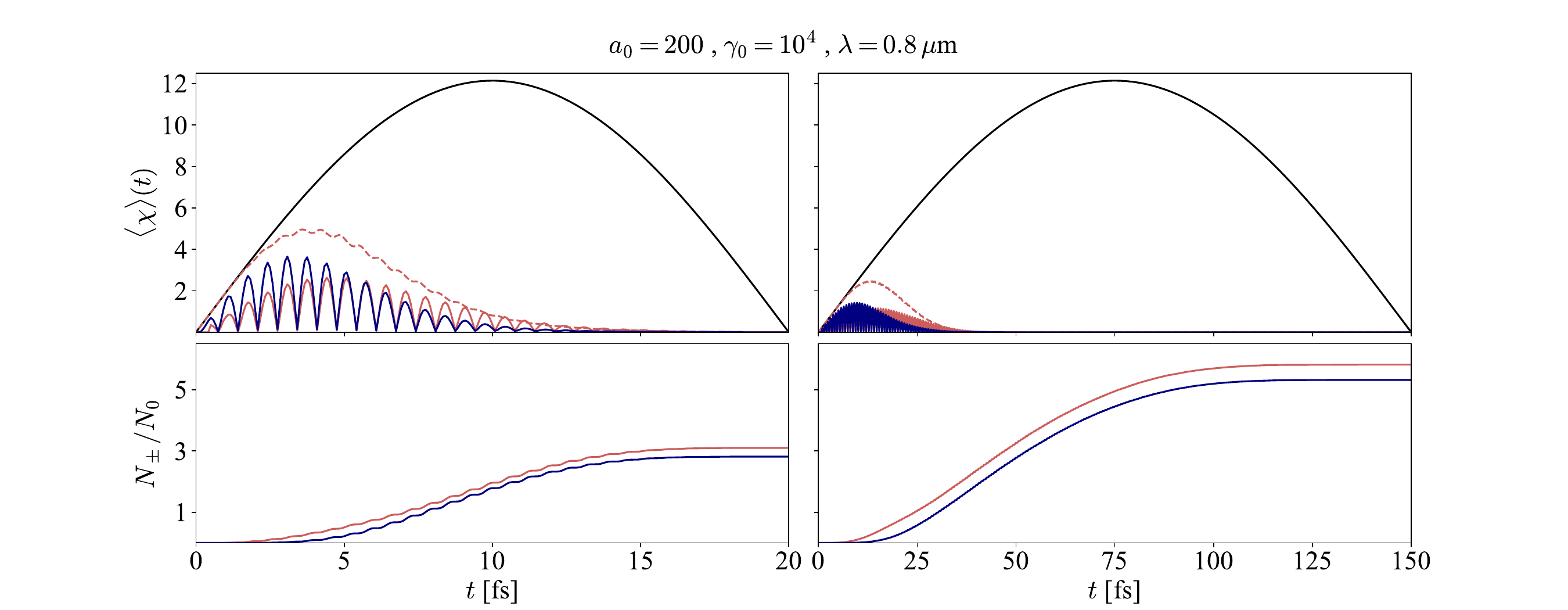}
    \caption{Simulated quantities as a function of time, for a short pulse (left panel, $\tau_p=20$ fs) and a long pulse (right panel $\tau_p=150$ fs). Incident energy of particles corresponds to $\gamma_0=10^{4}$, normalized intensity $a_0=200$ and  $\lambda=0.8 \, \mu$m for both simulations. Black line: maximum reachable quantum parameter $\chi$. Other solid lines: $\langle\chi\rangle$ of incident electrons or primary positrons (for photon seeding). Dashed red line: $\langle\chi\rangle$ of incident photons defined by $N_{\gamma}^{(0)}(t)\times \chi(t)$. The bottom panel shows the multiplicity as a function of time: blue (red) solid line for electron (photon) seeding.}
    \label{fig:long_short_pusle_chi}
\end{figure}

To illustrate the discussion, we show in Fig.~\ref{fig:long_short_pusle_chi} (top panel) the evolution of the quantum parameter of the incident electrons (in blue) or primary positrons (for photon seeding in red). Dashed red line represent $N_{\gamma}^{(0)}(t)\times \chi(t)$ in photon seeding case. On the bottom panel, we report the evolution of the multiplicity as a function of time in red for photons seeding and in blue for electrons seeding. In the case of a long pulse duration, the incident electrons dissipate their energy before reaching maximum intensity. As a result, the maximum average quantum parameters become less pronounced compared to a short-pulse case, and low-energy photon emission is favored. The competition between pair production rates over long and short periods can be succinctly characterized as follows: a set of low-energy photons interacting over a long duration versus a few high-energy photons interacting briefly.
The nonlinearities of the proposed equations prevent a direct resolution of this question. Simulations reveal growth in the number of pairs with pulse duration throughout the $a_0,\gamma_0$ parameter space explored see Fig.~\ref{fig:dependences_1D}(c). However, for some excessive parameters not shown here, we saw that the number of pairs can decrease with the pulse duration. From this analysis, it is important to understand that at the same intensity increasing the total pulse duration (so the total energy) is not always favorable to pair production. Also, it is important to note that for a longer pulse, pairs will also lose their energy by radiation and then have more chance to bounce back. The detectors have to be set with this consideration and an explicit study of the bounce effect is studied in \cite{qu2022collective}.

\subsection{Comparison to previous models}

\begin{figure}
    \includegraphics[width=1.\linewidth]{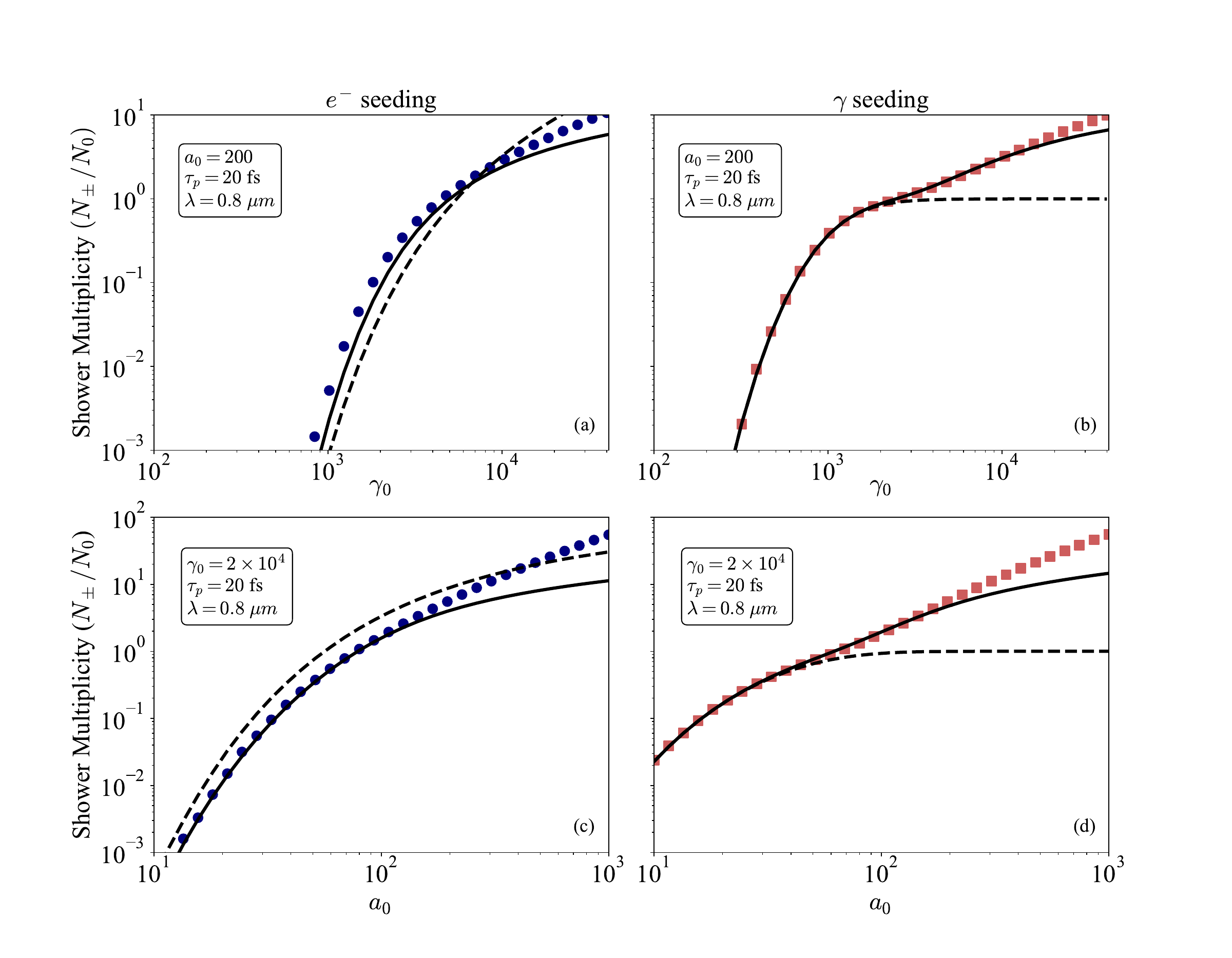}
    \caption{Shower multiplicities considering electron seeding (left panel) and photon seeding (right panel) as a function of incident particle normalized energy $\gamma_0$ considering $a_0=200$ and as a function of the laser field strength $a_0$ considering $\gamma_0=2\times10^4$. Red squares and blue dots are extracted from 1D3V PIC simulations and correspond to photon and electron seeding respectively. Black solid lines are obtained from our model (Eq. \eqref{eq:es:nbofpairs} for electron seeding and Eqs. \eqref{eq:ps:Npm1} and \eqref{eq:ps:Npm2} for photon seeding). Black dotted lines stand for the prediction of previous studies (\cite{blackburn2017scaling} in the cases of electron seeding and \cite{mercuri2021impact} and photon seeding).}
    \label{fig:comparison_1D}
\end{figure}

Calculations of the multiplicity in a similar configuration to the one considered here have been performed before by Blackburn et al. \cite{blackburn2017scaling} for the electron seeding case and by Mercuri-Baron et al. \cite{mercuri2021impact} for the photon seeding one. In this section, we compare our results with these previous models and further discuss the different hypotheses in use and their validity.

In Fig.~\ref{fig:comparison_1D}(a,b) we report the final number of pairs produced per incident particle as a function of the initial energy $\gamma_0$ and $a_0=200$ [panels (a) and (b) for electron- and photon seeding, respectively]. 
In Fig.~\ref{fig:comparison_1D}(c,d) we report the same quantity as a function of $a_0$ and fixed $\gamma_0=2\times10^4$ [panels (c) and (d) for electron- and photon seeding, respectively]. 
In all cases, the pulse parameters are $\tau_p=20$ fs and $\lambda=0.8$ $\mu$m. Red squares (photon seeding) and blue dots (electron seeding) are extracted from 1D3V PIC simulations. Our model predictions, Eq.~\eqref{eq:es:nbofpairs} (electron seeding) or Eqs.~\eqref{eq:ps:Npm1} and~\eqref{eq:ps:Npm2} (photon seeding), are plotted as solid lines. 

We first focus on electron seeding [panels (a) and (c)] and compare our predictions with those obtained using the model of Blackburn et al.~\cite{blackburn2017scaling}, shown as black dotted lines. In that paper, to propose a completely analytical model, the authors performed several approximations: i) The expression of the rate and not the real probability is used for pair creation. This induces a systematic error when the multiplicity $\gtrapprox 0.1$ (or more accurately when $\Wbw \simeq 1/\tau_p$), leading to an overestimate of the number of produced pairs that, at high probability, can even overcome the number of available photons. ii) The nonlinear Compton Scattering rate is asymptotically expanded around   $\gamma -\gamma_\gamma \ll 1$. This is a very good assumption, as verified in our simulations, when the highest-energy photons are the ones responsible for pair creation, which is the case in the low-$\chi_0$, low-multiplicity regime. However, as $\chi_0$ and the shower multiplicity increase, this approach fails in capturing the lower-energy photons that can contribute further to the pair production, and this assumption leads to underestimating the final number of final pairs.  iii) The nonlinear Compton Scattering rate is evaluated at the initial electron energy $\gamma_0$ and yet considering the time-dependent quantum parameter $\langle \chi \rangle$. This contrasts with our model (see Sec.~\ref{sec:model}) we consistently evaluate the rate of photon emission using the time-dependent average energy and quantum parameter. iv) Finally, the number of pairs is calculated heuristically integrating (over energies) a function built as the product of the nonlinear Breit-Wheeler rate by the emitted photon energy distribution. This approach leads to another systematic overestimate of the number of produced pairs. In our work, we rather use the exact form given by Eq.~\eqref{eq:es:nbofpairs} of this work. 

Thanks to these approximations, Blackburn et al. obtain a fully analytical expression for the number of pairs produced in a shower driven by a linearly polarised Gaussian laser. Their prediction appears in fair agreement with simulations shown in Figs. \ref{fig:comparison_1D}~(a) and~(c).

Our approach extends Blackburn's by alleviating certain hypotheses that are not well justified in the regime of intermediate quantum parameters $\chi_0$ and potentially large shower multiplicities relevant to multi-petawatt laser facilities. 
Doing so however requires a numerical integration of the main equations derived in Sec.~\ref{sec:model:es}, but also allows for a more general treatment, not limited to the case of linearly polarized Gaussian beam.\\

Let us now turn our attention to the case of photon seeding. In Figs.~\ref{fig:comparison_1D}(b) and~(d) we show the prediction of the shower multiplicity obtained from the model of Mercuri-Baron et al. \cite{mercuri2021impact} (black dotted lines). In this case, the multiplicity is estimated by considering exactly one generation, i.e. assuming the soft-shower regime. The model is equivalent to Eq.~eq\ref{eq:ps:Npm1} and as we can see gives a very good agreement with simulations until the multiplicity becomes larger than $1$.  Note that in the work of \cite{mercuri2021impact}, a fully analytical expression of Eq.~\ref{eq:ps:Npm1} is given for arbitrary geometry (see also Appendix~\ref{appendixC}). 

In this work, we improve on the model of Mercuri-Baron et al. by computing the number of pairs of the second generation. Doing so extends the model of photon-seeded shower to multiplicities larger than 1, see Figs.~\ref{fig:comparison_1D} (a)-(d), and allows to cover the whole range of parameters relevant to forthcoming facilities.\\

To conclude, the models developed in Sec.~\ref{sec:model}
improve on previous works by Blackburn, Mercuri-Baron, and co-workers, and extend our prediction capabilities to multiplicities of the order of or larger than unity, relevant for multi-petawatt laser facilities. At very large values of $\chi_0$, our models however underestimate the final number of pairs. This happens for multiplicities above typically 5, for which PIC simulations show that later generations of pairs and high-energy photons need to be accounted for. Such regimes of abundant pair production may not be easily accessible on near-feature facilities.


\section{Three-dimensional effects \& guidelines for future experiments}\label{sec:3d}

In this section, we generalize the model developed in Sec.~\ref{sec:model} to account for three dimensions. The model predictions are then benchmarked against 3D3V PIC simulations. Finally, predictions for pair multiplicity under forthcoming experimental conditions are discussed.

\subsection{Three-dimensional model}\label{sec:3D_model}

To construct our three-dimensional model, we consider the head-on collision between 
a linearly polarized Gaussian laser pulse with beam waist $w_0$ 
(radius at $1/e^2$ in intensity) propagating in the $+z$-direction 
(henceforth referred to as the longitudinal direction)
and an electron or gamma-photon beam propagating in the $-z$-direction. 
The latter (seed particle) beam has a characteristic longitudinal length $L_b$ and transverse size $w_b$
and is described, right before the collision, by its energy distribution function $dn_b/d\gamma$ which we parameterize as follows:
\begin{eqnarray}\label{eq:model3d:dnbdgamma0}
    \frac{d n_b}{d\gamma} = n_b\,f_b(x_0,y_0)\,g_b(z_0)\,h_b(\gamma)\,.
\end{eqnarray}
Here $(x_0,y_0,z_0)$ denotes the position of the seed particles before the collision, $n_b$ stands for the seed particle beam density, $f_b(x_0,y_0)$ and $g_b(z_0)$ the transverse and longitudinal beam density profile normalized such that\footnote{For a rectangular beam profile with transverse widths $w_{bx}$ and $w_{by}$ one thus has $w_b^2=w_{bx}\,w_{by}$. For a cylindrical beam with transverse radius $r_b$, one has $w_b^2=\pi r_b^2$.} $\int\!dx_0\,dy_0\,f_b = w_b^2$and $\int\!dz_0\,g_b = L_b$, and $h_b(\gamma)$ the beam energy distribution normalized such that $\int\!d\gamma\,h_b = 1$. The total number of particles in the beam then reads $N_b = n_b\,L_b\,w_b^2$.

The total number of pairs resulting from the collision can then be written in the form:
\begin{eqnarray}\label{eq:model3d:Ntot}
    N_\pm^{\rm 3D} = \int\!d\gamma\,\int\!dx_0\,dy_0\,dz_0\,\frac{dn_b}{d\gamma}\,M_{\rm traj}(\gamma;x_0,y_0,z_0)\,,
\end{eqnarray}
where $M_{\rm traj}(\gamma;x_0,y_0,z_0)$ measures the number of pairs produced by a single seed particle with energy $\gamma mc^2$
initially located at $(x_0,y_0,z_0)$ and considering its full trajectory across the laser pulse.

To compute Eq.~\eqref{eq:model3d:Ntot}, we now introduce some assumptions on both the laser pulse and seed particle beam.
As done for Sec.~\ref{sec:model}, we first assume that all (seed and produced) particles move at 
ultra-relativistic speed along straight lines (see Sec.~\ref{sec:model:hyp} for details) so that the trajectories over which $M_{\rm traj}$ is computed are simple. 
We further assume that the laser and seed particle beams are perfectly synchronized and aligned\footnote{That is, an ideal particle located at the center of the seeding beam and moving at the speed of light will reach the peak of the laser pulse at its focal plane.} and that the characteristic (longitudinal) length $L_b \equiv \int\!dz_0\,g_b(z_0)$ of the seed particle beam is smaller than or at most of the order of the laser beam Rayleigh length $z_R = \pi w_0^2/\lambda$ and pulse length $c\tau_p$. This ensures that all seed particles within the focal spot can experience the laser field at its maximum and makes the integration over $z_0$ in Eq.~\eqref{eq:model3d:Ntot} straightforward ($M_{\rm traj}$ can then be taken independent of $z_0$).
Under these assumptions, Eq.~\eqref{eq:model3d:Ntot} can be rewritten:
\begin{eqnarray}\label{eq:model3d:Ntot-bis}
    N_\pm^{3D} = n_b\,L_b\,\int\!d\gamma\,h_b(\gamma)\,\frac{d\sigma_{\rm tot}}{d\gamma} 
    \quad{\rm with}\quad 
    \frac{d\sigma_{\rm tot}}{d\gamma} \equiv \int\!dx_0\,dy_0\,f_b(x_0,y_0)\,M_{\rm traj}(\gamma;x_0,y_0)\,.
\end{eqnarray}

Considering that pair production occurs mainly within the focal volume, the laser electromagnetic field is well approximated by simply multiplying Eq.~\eqref{eq:Elaser} by a Gaussian dependency $\exp\left(-(x_0^2+y_0^2)/w_0^2\right)$. 
As a result, $M_{\rm traj} = N_\pm^{1D}/N_0$ is easily computed from the one-dimensional developed in Sec.~\ref{sec:model},
using $N_\pm^{1D} = N_\pm^{(1)}$ as given by Eq.~\eqref{eq:es:nbofpairs} for electron-seeded showers,
and $N_\pm^{1D} = (N_\pm^{(1)}+N_\pm^{(2)})$ with $N_\pm^{(1)}$ and $N_\pm^{(2)}$ given by Eqs.~\eqref{eq:ps:Npm1} and~\eqref{eq:ps:Npm2}, respectively, for photon-seeded showers.
Here, $N_\pm^{(1)}$ and $N_\pm^{(2)}$ are computed for all values of $x_0$ and $y_0$ using the laser field strength evaluated at $(x_0,y_0)$.

In the limit of a thin seed-particle beam, $w_b \ll w_0$, 
one can make the approximation $f_b(x_0,y_0) \simeq w_b^2\,\delta(x_0)\,\delta(y_0)$ in Eq.~\eqref{eq:model3d:Ntot-bis}.
The resulting number of produced pairs can then be simply extracted 
from the one-dimensional predictions of Sec.~\ref{sec:model}:
\begin{eqnarray}\label{eq:model3d:Ntot-ThinBeam}
    N_\pm^{3D} \xrightarrow{w_b \ll w_0} n_b\,L_b\,w_b^2\,\int\!d\gamma\,\left.\frac{N_\pm^{1D}}{N_0}\right\vert_{\gamma,x_0=y_0=0}\,.
\end{eqnarray}

In the limit of a broad (pancake-like) seed-particle beam, $w_b \gg w_0$,
one can take $f_b(x_0,y_0) \simeq 1$ in Eq.~\eqref{eq:model3d:Ntot-bis} and the number of produced pairs can be computed as:
\begin{eqnarray}\label{eq:model3d:Ntot-PancakeBeam}
    N_\pm^{3D} \xrightarrow{w_b \gg w_0} n_b\,L_b\,\sigma_{\rm tot} \quad {\rm with} \quad \sigma_{\rm tot} \equiv \int\!d\gamma\,h_b(\gamma)\,\frac{d\sigma_{\rm tot}}{d\gamma}\quad {\rm and} \quad \frac{d\sigma_{\rm tot}}{d\gamma} = \int\!dx_0\,dy_0\,\left.\frac{N_\pm^{1D}}{N_0}\right\vert_{\gamma;x_0,y_0}\,.
\end{eqnarray}
For this broad (pancake-like) seed-particle beam, we see that 
the electron beam is fully defined by its areal density $n_b\,L_b$
and the total number of produced pairs is proportional to an effective, total cross-section $\sigma_{\rm tot}$ (as introduced in Ref.~\cite{mercuri2021impact}). 
It is interesting to rewrite Eq.~\eqref{eq:model3d:Ntot-PancakeBeam} as $N_\pm^{3D} = n_b\,L_b\,w_0^2\,M_{\rm eff}$, where $n_b\,L_b\,w_0^2$ is the number of seed particles in the focal volume and $M_{\rm eff} \equiv \sigma_{\rm tot}/w_0^2$ thus
measures an effective multiplicity. For a Gaussian laser pulse, this effective multiplicity is independent of $w_0$.

In what follows, we benchmark this three-dimensional model against
self-consistent particle-in-cell simulations. As the case of a thin beam is a direct extrapolation of one-dimensional results, we rather focus our attention on the case of a pancake beam ($w_b \gg w_0$). In addition, we now focus our attention on the case of a monochromatic seed-particle beam [$h_b(\gamma) = \delta(\gamma-\gamma_0)$].

\subsection{3D3V PIC simulations}

\begin{figure}
    \includegraphics[width=0.9\linewidth]{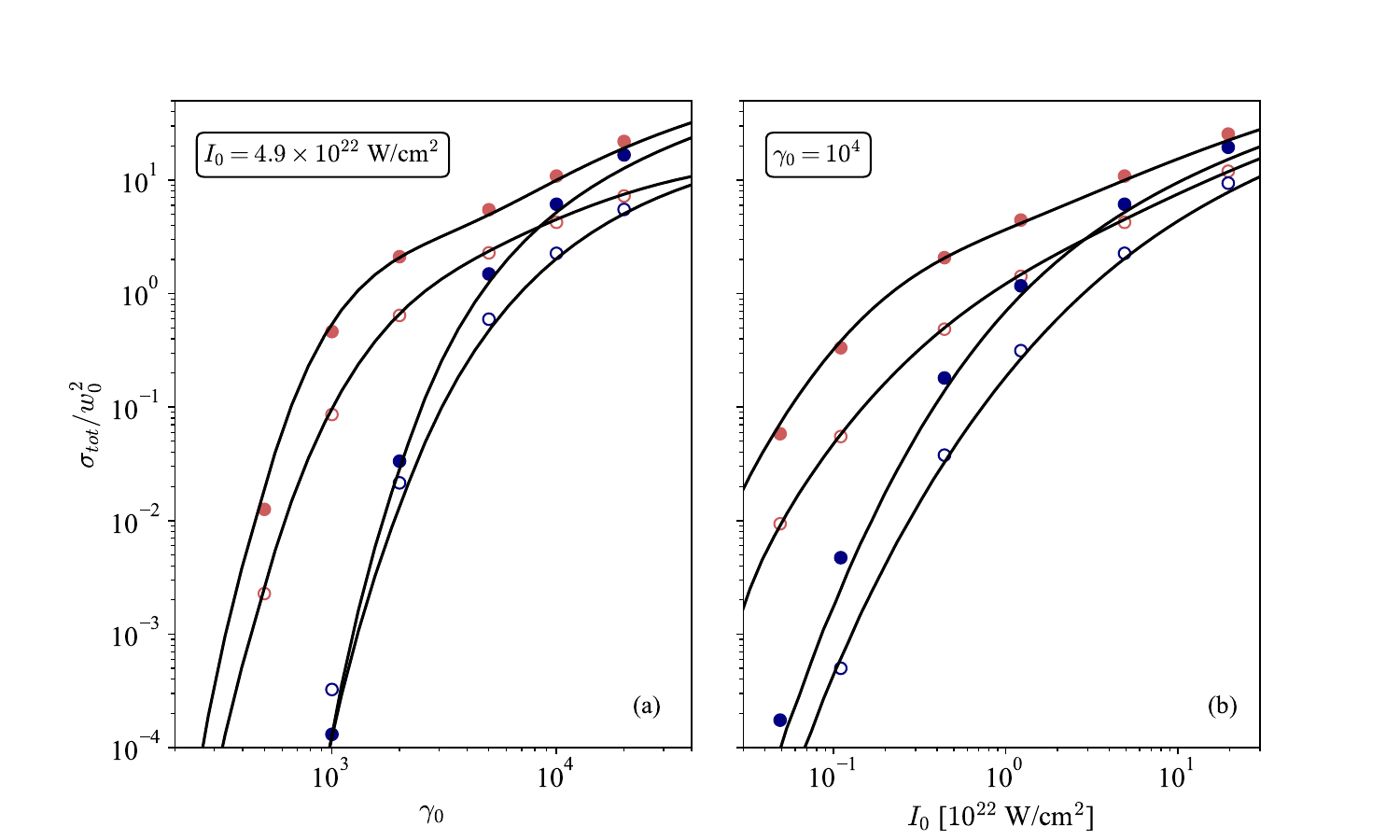}
    \caption{Total cross section obtained from 3D PIC simulations as a function of the incident particle normalized energy $\gamma_0$ (left panel) and the laser intensity $I_0$ (right panel). It is normalized to $w_0^2$ (with $w_0$ the beam waist) so that $\sigma_{\rm tot}/w_0^2$ measures an effective multiplicity. Empty circles correspond to conditions relevant for Apollon-like lasers ($\tau_p=20$ fs, $\lambda=0.8$ $\mu$m) and full circles to ELI-like lasers ($\tau_p=150$ fs, $\lambda=1.057$ $\mu$m). In blue: electron-seeded showers. In red: photon-seeded showers. Black lines correspond to the predictions from our model [Eq.~\eqref{eq:model3d:Ntot-PancakeBeam}]. } 
    \label{fig:3D_compare}
\end{figure}

\begin{figure}
    \includegraphics[width=0.8\linewidth]{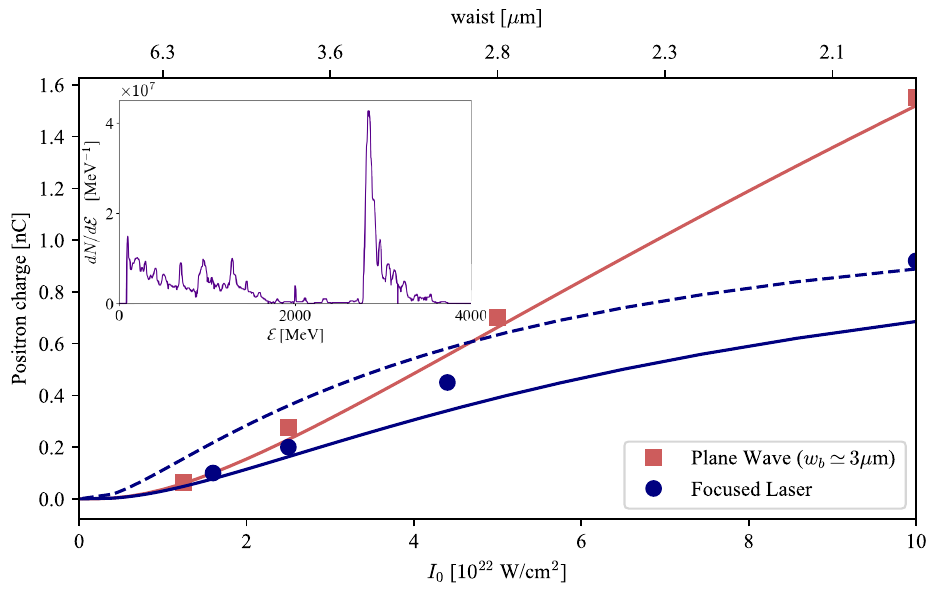}
    \caption{Final positron charge as a function of the maximal laser intensity for the beam of electron defined in \cite{lobet2017generation}. Red square correspond to the simulation of \cite{lobet2017generation} for Plane Wave laser. Blue dots correspond to the simulation results of \cite{lobet2017generation} for focused laser with the same duration  and energy but with various spot sizes. Red and blue solid lines are the predictions obtained from Eq.~\eqref{eq:model3d:Ntot} for these two different laser definitions. Blue dashed line correspond to the prediction using a pancake-like beam. In top left panel, the energy distribution of the incident electron from \cite{lobet2017generation}.} 
    \label{fig:lobet_compar}
\end{figure}

\begin{figure}
    \includegraphics[width=1\linewidth]{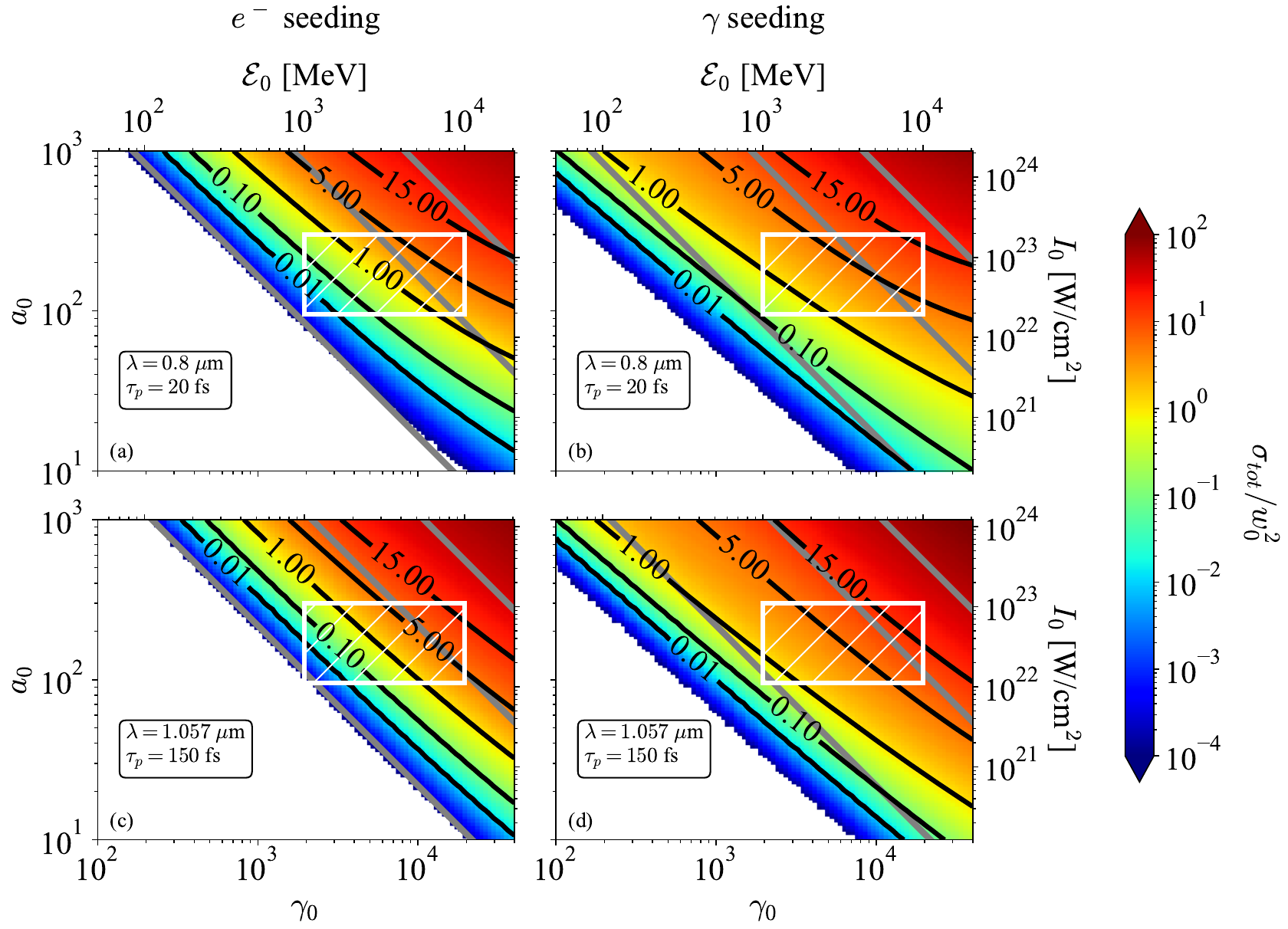}
    \caption{Theoretical cross-section normalised to $w_0^2$ (or effective multiplicity)from Eq.~\eqref{eq:model3d:Ntot-PancakeBeam} considering electron seeding (left panels) 
        and photon seeding (right panels) as a function of the laser field strength $a_0$ (equivalently intensity $I_0$) 
        and incident particle normalized energy $\gamma_0$ (equivalently $\mathcal{E}_0$). 
        For short pulse ($\tau_p=20$ fs) Ti:Sapphire lasers (top panels) and longer pulse ($\tau_p=150$ fs) Nd:glass (bottom panels). The white region corresponds to the region of interest for PW to multi-PW laser systems.
        Black solid lines report iso-contours $0.01,0.1,1,5,15$ and grey lines stand for $\chi_0 = 1$, $10$ and $50$ (from bottom-left to top-right). Note that the colormap saturates for $\sigma_{tot}/w_0^2<10^{-4}$. } 
    \label{fig:3D_sigma}
\end{figure}

In what follows we present 3D3V PIC simulations performed with Smilei. Each simulation reproduces a volume of $14\lambda \times 14 \lambda \times (c\tau_p/\lambda+1)\lambda$ (in the $x$, $y$ and $z$ directions respectively) with spatial resolution $\Delta x = \Delta y = ~\lambda/16$ and $\Delta z = ~\lambda/32$. Time resolution is set to $c \Delta t = ~\Delta z/2$. 
The laser propagates in the $z$-direction and collides head-on with a counter-propagating beam of electrons or photons with finite longitudinal extension $L_b=~\lambda/8$ and an infinite transverse extension (periodic boundary conditions are used in the transverse direction). Each cell of the beam contains $4$ macro particles for a total of $802816$ (seed) particles. The initial position of the particle beam is set so that if a particle at the center of the particle beam was always moving at $c$, it would reach the peak intensity at the beam focus. The duration of the simulation is longer than the interaction time of the laser and the particles. All results presented hereafter have been extracted at the end of the simulations. 

A set of simulations was performed considering the two types of laser facilities previously discussed, i.e. short Ti:Sapphire laser systems such as Apollon ($\lambda=0.8$ $\mu$m and $\tau_p=20$ fs), and long Ne:Glass laser systems such as the ELI Beamlines A4 ATON ($\lambda=1.057$ $\mu$m and $\tau_p=150$ fs). To compare the results on both types of facilities, results are presented as a function of the laser intensity (not field strength $a_0$). In addition, simulations have been performed using a fixed focusing aperture $f/3$.

In Fig.~\ref{fig:3D_compare}, the resulting total cross sections (dots), in units of $w_0^2$, are compared to those from the numerical integration of Eq.~\eqref{eq:model3d:Ntot-PancakeBeam} (lines), as a function of $I_0$ and $\gamma_0$. Excellent agreement is observed for the range of parameters considered.
Since the effect of finite transverse size is that many particles will interact with an intensity lower than the peak value $a_0$, the approximation of considering only the first generation holds over a wider range of parameters in the three-dimensional case than in the one-dimensional limit. If we compare pulses with different duration and same field peak amplitude (here $I_0=4.9\times 10^{22}$ W/cm$^2$ i.e. $a_0=200$ at $\lambda=1.057$ $\mu$m) it is interesting to notice that in the case of electron seeding, for the lowest electron energies we obtain a slightly larger cross-section for $\tau_p=20$ fs than $\tau_p=150$ fs. As discussed before, electrons lose more energy before reaching the peak of intensity in a long pulse, resulting in nonlinear Compton Scattering emitted photons with lower energy. This means that the number of pairs is not always a growing function of the pulse duration. 

It is also interesting to compare  the number of created pairs predicted by our model with previous self-consistent full-3D PIC simulations performed in Ref.~\cite{lobet2017generation}. As discussed in details below, the comparison shows excellent agreement and  allows for quickly estimating the order of magnitude of expected pairs. The comparison is shown in Fig.~\ref{fig:lobet_compar}. The pulse parameters are $\tau_p=15$ fs, $\lambda=0.8$ $\mu$m and the waist is variable in the different simulations (from 2 to 5 $\mu$m corresponding to peak intensities from $10^{23}$ to $2.5\times 10^{22}$ W/cm$^2$). The incident beam is generated by laser wake-field acceleration and contains approximately $1$ nC of $\approx 3$ GeV electrons. Considering only these particles, the density is $\approx 10$ C/cm$^3$ with a longitudinal size of $12$ $\mu$m. A first estimate can be obtained using Fig.~\ref{fig:3D_sigma} that reports our results with very similar laser parameters: $\sigma_{tot}/w_0^2 \approx 2$.
The positron charge is then  $e\times\sigma_{tot}/w_0^2\times n_b L_b w_0^2 \approx 1 $ nC, in excellent agreement with the $\approx 0.9$ nC obtained by full-3D PIC simulations in Ref.~\cite{lobet2017generation}.

To obtain a more precise estimation, we extracted from their work the full energy spectrum of the incident electrons (shown in the inset panel of Fig.\ref{fig:lobet_compar}) and 
performed the full calculation using Eq. \eqref{eq:model3d:Ntot}. In Fig.~\ref{fig:lobet_compar}, the red color stand for laser plane wave simulations with different intensities; the squares show the result of the 3D PIC simulation of \cite{lobet2017generation} while the red line is obtained with Eq.~\eqref{eq:model3d:Ntot}.  In this case the laser intensity does not depend on space and the parameters needed are the energy spectrum and the number of initial electrons. The calculation can then be performed in one dimension, and related to the multiplicity discussed in Sec.~\ref{sec:1dpic:Multip_and_plot}. Notice that, in this case, the number of interacting electrons is kept constant and equal to the total number of electrons in the beam in the 3D simulations. The agreement with our calculation is excellent. 
Blue points are the results presented in Ref.~\cite{lobet2017generation} for a $60$-J focused laser pulse with various spot sizes. The solid blue line corresponds to our prediction using Eq. \eqref{eq:model3d:Ntot} calculated with the full spatial distribution of the incident electron beam. We defined it as a cube of size $4$ $\mu$m, $2$ $\mu$m and $12$ $\mu$m along and normal to the laser polarization and along the propagation direction, respectively. The cube is aligned with the focal spot. In this prediction, we neglected diffraction, meaning that the longitudinal geometry is not taken into account, but we still predict the final number of pairs within less than a factor $2$. The dashed blue line shows our prediction for a pancake-like beam Eq.~\eqref{eq:model3d:Ntot-PancakeBeam}, considering a constant particle density (the number of interacting electrons increases with the waist).
The trend is very similar to the case when the interaction is limited to the central spot, and as expected, for a waist of order or lower than the beam transverse size ($\lessapprox 3$ $\mu$m) the results are in better agreement with the full 3D PIC simulation.  Notice however that, even if the results of the pancake beam show a higher charge at large waist than the other two cases, a  direct comparison in this range is not appropriate, since the total number of interacting electrons is not the same. 

Finally, to guide future experiments we show in Fig.~\ref{fig:3D_sigma} the theoretical total cross-section $\sigma_{tot}=N_{pair}/(L_bn_b)$ normalized to $w_0^2$ for a pancake-like beam as a function of laser field intensity and initial incident energy for pulses like ELI and Apollon.  A generalization to other situations can be found  in \cite{amaro2021optimal}. In our case the cross section is proportional to  $w_0^2$ and we present the value $\sigma_{tot}/w_0^2$ corresponding to an effective multiplicity.  Thus, the number of created pairs after the interaction can be easily obtained as the number of incident particles initially in the waist times the effective multiplicity. In other words, a good estimation of the final number of pairs created can be calculated by: $N_\pm=(\sigma_{tot}(\gamma_0,a_0,\tau_p)/w_0^2)\times n_bL_b w_0^2$. If the initial beam is spread in energy, this formula can simply be multiplied by the initial energy spectrum and integrated over energies.

\section{Conclusions}\label{sec:conclusion}

In conclusion, we have developed a semi-analytical model that predicts the multiplicity of pair showers emerging from the head-on collision of high-energy seed particles (electrons or photons) with an ultra-high-intensity laser pulse. 
This model builds on previous works by Blackburn et al.~\cite{blackburn2017scaling} and Mercuri-Baron et al.~\cite{mercuri2021impact} and extends their domain of validity to the regime of higher multiplicities and conditions relevant to forthcoming multi-petawatt laser facilities. 
The model follows the evolution of successive generations of high-energy photons and electron-positron pairs. 
It is shown that under conditions of interest for this study, that correspond nonetheless to a large range of laser field strengths and seed particle energies, only the first generations contribute to the shower while large multiplicities, in the range 1 to 15, can still be achieved. 
This result is confirmed by both one-dimensional and three-dimensional PIC simulations, and the model is used to draw guidelines for future experiments.

The difference between electron- and photon-seeded showers is clarified.
In the regime of moderate quantum parameter, $\chi_0 \lesssim 1$, small multiplicities are expected and photon seeding is significantly more favorable than electron seeding (as far as we are concerned with multiplicity and not absolute numbers of produced particles). 
At higher quantum parameters, and for higher multiplicities, both seeding strategies lead to equivalent multiplicities so that 
electron seeding can be expected to lead to a larger (absolute) number of produced pairs.

The effect of pulse duration (for a given peak intensity) is also discussed. 
At high intensities, relevant to forthcoming experiments at multi-petawatt laser facilities, pair production is weakly dependent on the interaction time. For long pulses, the primary particles mainly interact or lose energy in the initial part of the pulse and do not attain the maximum reachable quantum parameter. Contrary to what could be naively expected, we find that the number of produced pairs does not increase linearly with the pulse duration (nor energy). In practical terms, it means that, considering a similar maximum intensity for the laser pulse, the multiplicity of showers using long Nd:Glass laser is only marginally increased compared to what can be achieved using a short Ti:Sapphire laser.

Last we want to point out that, while this work focused on head-on collisions of high-energy electron or photon beams against linearly-polarised lasers with Gaussian (in space) and $\sin^2$ (in time) beam profile, our model — and in particular Eqs.~\eqref{eq:es:gamma_moy_vs_time}, \eqref{eq:es:nbofpairs}, \eqref{eq:ps:gamma1}, \eqref{eq:ps:Npm1} and \eqref{eq:ps:Npm2} — can be used considering arbitrary polarizations, spatio-temporal laser beam shapes and collision angles. 
Furthermore, while we discussed mainly monoenergetic and very thin (in the longitudinal direction) particle beams, extension to arbitrary energy and spatial distribution can be easily obtained.

\section*{Acknowledgements}
The authors are grateful to E. Gelfer and S. Weber for fruitful discussions. This work used the open-source PIC code SMILEI, the authors are grateful to all SMILEI contributors and to the SMILEI-dev team for its support. Simulations were performed on the Irene-Joliot-Curie machine hosted at TGCC, France, using High-Performance Computing resources from GENCI-TGCC (Grant No. A0030507678). This work received financial support from the French state agency \textit{Agence Nationale de la Recherche}, in the context of the \textit{Investissements d'Avenir} program (reference ANR-18-EURE-0014). A.A.M. was supported by Sorbonne Universit\'e in the framework of the Initiative Physique des Infinis (IDEX SUPER). The work of T.G. and M.V. was supported by FCT grants: CEECIND/04050/2021, CEECIND/01906/2018 and PTDC/FISPLA/3800/2021.

\appendix

\section{Governing equations for the full distribution functions}\label{appendixA}

Placing ourselves within the locally-constant crossed field approximation (LCFA), neglecting particle acceleration in the laser field and any spatial non-uniformities, one can cast the cascade equations in the form:
\begin{eqnarray}
    \label{eq:equation_coupled_el} \partial_t f_\pm(\gamma,t)&=&\int_0^\infty d \gamma_\gamma w_\chi(\gamma+\gamma_\gamma,\gamma_\gamma)f_\pm(\gamma+\gamma_\gamma,t) -f_\pm(\gamma,t)\int_0^\infty d \gamma_\gamma w_\chi(\gamma,\gamma_\gamma)+\int_0^\infty d \gamma_\gamma \overline{w}_{\chi_\gamma}(\gamma_\gamma,\gamma)f_\gamma(\gamma_\gamma,t)\,,\quad\\
    \label{eq:equation_coupled_pho} \partial_t f_\gamma(\gamma_\gamma,t)&=&\int_1^\infty d \gamma w_{\chi}(\gamma,\gamma_\gamma)f_+(\gamma,t)+\int_1^\infty d \gamma w_{\chi}(\gamma,\gamma_\gamma)f_-(\gamma,t) -\int_1^\infty d \gamma \overline{w}_{\chi_\gamma}(\gamma,\gamma_\gamma)f_\gamma(\gamma_\gamma,t)\,.
\end{eqnarray}
Equation~\eqref{eq:equation_coupled_el} defines the dynamics of the energy distribution of electrons ($f_-$) and positrons ($f_+$). The first two terms in the equation right-hand-side describe the effect of high-energy photon emission via nonlinear inverse Compton scattering on the electron/positron dynamics (radiation reaction).
The last term describes the creation of new pairs by \textit{nonlinear Breit Wheeler} and depends on photon distribution only. Finally, the evolution of the photon energy distribution ($f_\gamma$) is given by Eq.~\eqref{eq:equation_coupled_pho}. The first two terms account for the emission of new photons by electrons and positrons, respectively, and the last term corresponds to the decay of photons into new pairs.

The energy distribution functions $f_-$, $f_+$ and $f_\gamma$ as described by Eqs.~\eqref{eq:equation_coupled_el} and \eqref{eq:equation_coupled_pho} account for all electrons, positrons and $\gamma$ photons independently of their origin. 
In this work, we have shown that it can be interesting to split each population in successive generations that may be treated separately. 
We define $f_-^{(n)}(\gamma,t)$ and $f_+^{(n)}(\gamma,t)$ the distribution function of electrons and positrons created by photons of the $(n-1)^{th}$ generation. Similarly, photons of the $n^{th}$ generation are those created by the $n^{th}$ generation of electrons and positrons, and their distribution function is denoted $f_\gamma^{(n)}(\gamma_\gamma,t)$. From this definition, it is clear that the full distribution function of photons $f_\gamma$ electron $f_-$ and positron $f_+$ can be reconstructed summing over all particle generations: $f_s = \sum_n f_s^{(n)}$ with $s=\pm,\gamma$. 

The equation evolution governing the $n^{th}$ generation distribution functions read:
\begin{eqnarray}
    \nonumber \partial_t f_\pm^{(n)}(\gamma,t) &=& \int_0^{\infty}\!\! d\gamma_\gamma\, w_\chi(\gamma+\gamma_\gamma,\gamma_\gamma)\,f_\pm^{(n)}(\gamma+\gamma_\gamma,t) \nonumber\\
    &-& \int_0^{\infty}\!\! d\gamma_\gamma\, w_\chi(\gamma,\gamma_\gamma)\,f_\pm^{(n)}(\gamma,t) \nonumber\\
    &+&\int_0^{\infty}\!\! d\gamma_\gamma\, \overline{w}_{\chi_\gamma}(\gamma_\gamma,\gamma)\,f_\gamma^{(n-1)}(\gamma_\gamma,t)\,, \label{eq:es:fpn}\\
    \nonumber \partial_t f_\gamma^{(n)}(\gamma_\gamma,t) &=& \int_1^{\infty}\!\! d\gamma\, w_\chi(\gamma,\gamma_\gamma)\,f_-^{(n)}(\gamma,t)\\
    &+&\int_1^{\infty}\!\! d\gamma\, w_\chi(\gamma,\gamma_\gamma)\,f_+^{(n)}(\gamma,t) \nonumber\\
    \label{eq:es:fgn}&-& \int_1^{\infty}\!\! d\gamma\, \overline{w}_{\chi_\gamma}(\gamma_\gamma,\gamma)\,f_\gamma^{(n)}(\gamma_\gamma,t)\,.
\end{eqnarray}
A general solution of Eq.~\eqref{eq:es:fgn} can be obtained using the variation of parameter method, which gives:
\begin{eqnarray}\label{eq:fGn}
    f_\gamma^{(n)}(\gamma_\gamma,t)=A(\gamma_\gamma,t)\,\exp\left(-\int_0^t\!d\tau\,\Wbw(\gamma_\gamma,\chi_\gamma(\tau))\right)\,
\end{eqnarray}
with
\begin{eqnarray}
    A(\gamma_\gamma,t) = f_\gamma^{(n)}(\gamma_\gamma,t=0)+\int_1^\infty \!d\gamma \int_0^t\!dt'\, \left[f_-^{(n)}(\gamma,t)+f_+^{(n)}(\gamma,t)\right]w_\chi\!\left(\gamma,\gamma_\gamma\right)\,
    \exp\bigg(\int_0^{t'}\!d\tau\,\Wbw(\gamma_\gamma,\chi_\gamma(\tau))\bigg) \,.
\end{eqnarray}
Note that the term $f_\gamma^{(n)}(\gamma_\gamma,t=0)$ is non-zero only for generation $n=0$ in the case of a photon-seeded shower.

Integrating Eq.~\eqref{eq:es:fpn} over all charged particle energy ($\gamma$) and interaction time leads the number of pairs generated at the $(n+1)^{th}$ generation:
\begin{eqnarray}\label{eq:partialNn}
    N_\pm^{(n+1)}(t) = \int_0^\infty\!d\gamma_\gamma\,\int_0^t\!dt'\,\Wbw(\gamma_\gamma,\chi_\gamma(t'))\,f_\gamma^{(n)}(\gamma_\gamma,t')\,.
\end{eqnarray}
Injecting Eq.~\eqref{eq:fGn} into \eqref{eq:partialNn} and integrating by part gives:
\begin{eqnarray}\label{eq:Nn}
    N_\pm^{(n+1)}(t) &=& \,\int_0^\infty\!d\gamma_\gamma\,
    \int_0^t\!dt'\, \int_1^\infty \!d\gamma \left[f_{-}^{(n)}(\gamma,t)+f_{+}^{(n)} (\gamma,t)\right]w_\chi(\gamma,\gamma_\gamma)\,
    \left[1-\exp\left(-\int_{t'}^t\!d\tau\,\Wbw(\gamma_\gamma,\chi_\gamma(\tau))\right)\right]\ \nonumber \\
    &+& \int_0^\infty\!d\gamma_\gamma\, f_\gamma^{(n)}(\gamma_\gamma,t=0)\left[1-\exp\left(-\int_{0}^t\!d\tau\,\Wbw(\gamma_\gamma,\chi_\gamma(\tau))\right)\right]\,.
\end{eqnarray}

The first term of this general expression can be intuited by considering that the creation of pairs at a given time corresponds to the probability to create a photon multiplied by the probability of this photon to decay during an interval of time equal to the difference of the given time and the photon creation time. This is expressed as a convolution of these two probabilities weighted by the distribution function of the parent charged particle. 

The second term takes into account the generation of pairs by the initial ($n^{th}$) photons but is non-zero only for the particular case of photon seeding.  

Let us finally note that, summing Eq.~\eqref{eq:Nn} over all successive generations, we obtain the total number of pairs as a function of the photon, electron, and positron distribution functions.

\section{Photon emission and pair production rates}\label{appendixB}

The (energy) differential rates Eqs.~\eqref{eq:w-nics} and~\eqref{eq:w-bw} are expressed in terms of the functions:
\begin{eqnarray}
    \label{eq:F-nics} F(\chi,\xi) &=& \frac{1}{\pi\sqrt{3}}\,\left[\frac{\xi^2}{1-\xi}\,K_{2/3}(\nu) + \int_\nu^{\infty}K_{5/3}(y)dy\right]\,,\\
    \label{eq:G-bw} G(\chi_\gamma,\zeta) &=& \frac{1}{\pi\sqrt{3}}\,\left[\frac{1}{\zeta(1-\zeta)}\,K_{2/3}(\mu) - \int_\mu^{\infty}K_{5/3}(y)dy\right]\,,
\end{eqnarray}
with $\xi=\gamma_\gamma/\gamma$, $\nu=2\xi/[3\chi(1-\xi)]$, $\zeta=\gamma/\gamma_\gamma$ and $\mu=2/[3\chi_\gamma\zeta(1-\zeta)]$,
and where $K_n(x)$ denotes the modified Bessel function of the second kind.\\

The (energy-integrated) rates of photon emission and pair production, Eqs.~\eqref{eq:W-nics} and~\eqref{eq:W-bw}, 
are expressed in terms of the functions:
\begin{eqnarray}
    \label{eq:a-nics} a(\chi)\!&=&\!\int_0^1 d\xi\,F(\chi,\xi) \longrightarrow 
    \begin{cases}
        1.44\,\chi, & {\rm for}~\chi \ll 1\\
        1.46\,\chi^{2/3}, & {\rm for}~\chi \gg 1
    \end{cases} \,,\\
    \label{eq:b-bw} b(\chi_\gamma)\!&=&\!\int_0^1 d\xi\,G(\chi_\gamma,\xi) \longrightarrow
    \begin{cases}
        0.23\,\chi_\gamma\,\exp\big(\!-8/(3\chi_\gamma)\big) & {\rm for}~\chi_\gamma \ll 1\\
        0.38\,\chi_\gamma^{2/3}, & {\rm for}~\chi_\gamma \gg 1
    \end{cases} \,.
\end{eqnarray}
To compute $b(\chi_\gamma)$ in our model, we use the approximate form proposed in Ref.~\cite{mercuri2021impact}
(accurate within 1\%):
\begin{eqnarray}
    b(\chi_\gamma) \simeq 0.161\,\frac{K_{1/3}^2\big(4/(3\chi_\gamma)\big)}{1-0.172/(1+0.295 \chi_\gamma^{2/3})}\,.
\end{eqnarray}

\section{Number of primary pairs in photon-seeded cascades}\label{appendixC}

In Ref.~\cite{mercuri2021impact}, the authors derive an
analytical approximation for the number of primary pairs created in a photon-seeded shower. Considering head-on collision with a linearly polarized light pulse described by Eq.~\eqref{eq:Elaser}, it reads:
\begin{eqnarray}\label{eq:ps:Npm1-mercuri}
    N_\pm^{(1)}(t) = N_0\,\left[1 - \exp\left(-R_n t \right)\right]\,,
\end{eqnarray}
with
\begin{eqnarray}
    R_n= \frac{1}{n}\sum_{l=1}^n R_l \,,
\end{eqnarray}
$n=\lfloor 2t\omega/\pi \rfloor$ ($\lfloor \, \rfloor$ denotes the integer part) denotes the number of half optical cycles a seed photon has experienced at a given time $t$, and:
\begin{eqnarray}
     R_l=\frac{3\alpha}{2\tau_c \gamma_0} b(\chi_l)\,\min\!\big\{F(s(\chi_l)),0.713\big\} \mbox{ with }
        F(s) = \sqrt{2/\pi}\,s\, \mbox{erf}(\pi\sqrt{2}/(4s))\,.
\end{eqnarray}
Here $\chi_l=\chi_0\sin[(1+2l)\pi^2/(4\omega \tau_p)]$ 
is the quantum parameter of a photon seeing the $l^{th}$ field extremum, computed considering a sin$^2$ intensity envelope [see Eq.~\eqref{eq:Elaser}], $\mbox{erf(x)}=(2/\sqrt{\pi})\int_0^x\!dy\,e^{-y^2}$ is the error function and $s(\chi)=\sqrt{b(\chi)/[\chi b'(\chi)]}$ with $b(\chi)$ defined in Appendix~\ref{appendixB} and $b'(\chi)$ its derivative.


%

\end{document}